\documentstyle[12pt,uuthesis,pks]{thesis}
\def \eps {\epsilon}
\def \th {\theta}
\def \nonu {\nonumber}
\def \ni {\noindent}
\def \cl {\centerline}
\def \c {\cite}
\begin{document}
\title{Study of the properties of dense nuclear matter and
application to some astrophysical systems}
\author{PRADIP KUMAR SAHU}
\guide{Professor L. Satpathy}
\coguide{Professor B. Datta}
\copyrightfalse
\figurespagefalse
\tablespagefalse
\beforepreface
\newpage
\begin{figure}
\vskip 1.5in
\cl{\LARGE{\underline{\it TO MY PARENTS}}}
\end{figure}
\prefacesection{Acknowledgements}
\hspace {0.3in} I express my gratitude to Professor L. Satpathy
and Professor B. Datta for their invaluable guidance and
encouragement. My association with them
during the investigation of the present thesis has been a rich
source of knowledge for me.
\par
I am very much thankful to Professor V. S. Ramamurthy, Director,
Institute of Physics, Bhubaneswar for his sincere encouragement and advice.
I am also grateful to Professor Ramnath Cowsik, Director, Indian
Institute of Astrophysics, Bangalore for the academic facilities
extended to me during various stages of my work.
\par
I would also like to acknowledge Professor (Mrs.) M. Satpathy, who is
responsible for introducing me to Nuclear Physics as well as
Nuclear Astrophysics.
\par
I am grateful to Professor S. C. Phatak for valuable research
collaboration with him and to Professors A. Ansari, A. Abbas, C.
R. Praharaj, H. C. Padhi, J. Maharana, B. B. Deo, N. Barik, L.
P. Singh, L. Maharana, Dr. K. Maharana and Dr. Subhendra Mohanty for
many useful discussions.
\par
I have greatly benefited from discussions and collaboration with
Dr. R. Basu of IMSc., Madras, Professors S. Hasan of IIA,
Bangalore, A. Goyal, J. D. Anand and Dr. B. C. Parija of Delhi
University, Delhi and V. S. Uma Maheswari, Dr. Prativa Sarangi and
Sanjay K. Ghosh of Institute of Physics, Bhubaneswar.
\par
It is a great pleasure for me to thank my friends Debanand Sa,
Sarira, Ananya, Sekar, Suresh K. Patra, Bijay, Mustafa, Satyam,
Prafulla, Suresh, Pratik, Giri, Supriya, Pijush, Subodh, Saroj,
Sundarvel, Haranath, Dutta, Kuri, Bipin, Nanda, Nandi, Umesh, Ravi,
Yadav, Manidipa, Sutapa and others from Institute of Physics,
Bhubaneswar and Arun V. Thampan, Pandey, Reddy, Sujan,
Annapurni, Uma, Ranganath, Krishna, Sivarani, Dilip and others from
Indian Institute of Astrophysics, Bangalore for their help.
\par
Finally, I have the great pleasure in expressing my thanks
to  my parents, brother Dilip and friends Sabyasachi, Saroj,
Sanjay, Ramakanta, Ramachandra, Rabaneswar, Mahendra and
Duryodhan for their valuable help and encouragement. My special
thanks to Uma Maheswari for her help, encouragement and discussions
during completion of this work.
\vskip 0.5 in
\ni Date: \hspace{4.0in} ( Pradip Kumar Sahu)
\prefacesection {List of Publications}
\begin{enumerate}
\bigskip
\item Eigenfrequencies of radial pulsations of strange quark
stars: by B. Datta, {\bf P.K. Sahu}, J.D. Anand and A. Goyal:
{\it Phys. Lett. B} {\bf 283} (1992) 313.
\item High density matter in the chiral sigma model: by {\bf
P.K. Sahu}, R. Basu and B. Datta: {\it Astrophys. J.} {\bf 416} (1993) 267.
\item Nuclear matter in the derivative scalar coupling
model: Energy per nucleon and finite temperature equation of state:
by B. Datta and {\bf P.K. Sahu}: {\it Phys. Lett. B} {\bf 318} (1993) 277.
\item Masses of multiquark droplets: by L. Satpathy, {\bf
P.K. Sahu} and V.S. Uma Maheswari: {\it Phys. Rev. D} {\bf 49} (1994)
4642.
\item Neutrino emissivity of non-equilibrium quark stars
matter: by {\bf P.K. Sahu}: {\it Int. Jour. Mod. Phys. E} {\bf 2} (1993) 647.
\item On the limitations of neutrino emissivity formula of
Iwamoto : by S.K.Ghosh, S.C. Phatak and {\bf P.K. Sahu}: {\it Mod.
Phys. Lett. A} (1994) (in press).
\item Quasi-molecular states in $^{12}C + ^{16}O$ and $ ^{16}O
+ ^{16}O$ systems: by L. Satpathy, {\bf P.K. Sahu} and
P.Sarangi: {\it J. Phys. G: Nucl. Part. Phys.} {\bf 18} (1992) 1793.
\item Quark stars in chiral colour dielectric model: by S.K.
Ghosh and {\bf P.K. Sahu.}: {\it Int. Jour. Mod. Phys. E} {\bf 2} (1993) 575.
\item Neutrino emissivity of degenerate diquark star matter:
by {\bf P.K. Sahu}:  {\it Mod. Phys. Lett. A} {\bf 8} (1993) 3435.
\end{enumerate}
\afterpreface
\newpage
\setcounter{equation}{0}
\chapter{Introduction}
\hspace {0.3in} In ordinary stars, energy is lost through
electromagnetic radiation
from the surface. In order to accomplish this, a gradient in the
temperature will be set up. The centre of the star will be
relatively hot and the surface layers relatively cool. As energy is
radiated away from the star, it shrinks just sufficient to provide
the energy loss, and an equal amount gets added to the
internally stored energy.
\par
The contraction of the star leads to a continual heating of the
stellar interior. A time will come when the temperature at the centre
will rise to $\sim 10^7~K$. At this stage, a new source of energy
will appear in the interior : fusion of hydrogen nuclei into helium
nuclei. Nuclear fusion releases enormous amounts of energy, which can
prevent the star's gravitational collapse. The star attains
hydrostatic equilibrium, and remains stable at a point on main
sequence in the Herztsprung--Russel diagram, determined largely
by its mass. The duration of this phase
is very long (billions of years). That is why most stars that we see
in the sky are main sequence stars.
\par
In the above process, eventually a time will come when the supply of
hydrogen in the star gets depleted. This leads to the following
sequence of events : (1) a drop in the energy production, (2)
collapse of the star and a consequent rise of temperature in the
core of the star (which is now mostly helium) and (3) fusion of
helium to carbon and
oxygen nuclei. This will lead to a second relatively stable phase of
the star's life. However, unlike the first stage, a qualitatively
new phenomenon now takes place. The energy release during the second
phase of core collapse will not only lead to core heating but, in
addition, cause an expansion of the outer layers of the star. As a
result, the star gets a bloated shape (expected radius $\sim 100$
million $km$), and the temperature of the outer layers drops. This is
called a red giant. The helium burning phase is expected from theory
to last several million years.
\par
The depletion of helium again leads to a core contraction and a
consequent rise in temperature. The further evolution of the
star depends on the initial mass of the star.
\par
According to stellar evolution theory, for stars whose initial mass
is less than about 6 $M_{\odot}$ ($M_{\odot}$ = solar mass), the
rise in temperature is insufficient to start the next fusion cycle
(carbon $\longrightarrow$ neon). So, the collapse continues, and the
energy released eventually blows off the remaining outer layers
of the star. This produces a remnant stellar core and an expanding envelope.
Such objects are called planetary nebulae, and some of these have
been observed in the sky. The duration of this evolutionary phase is
relatively short, about tens of thousands of years.
\par
The remnant stellar core does not undergo gravitational contraction
indefinitely. To understand this, one has to recall the Pauli
exclusion principle in quantum mechanics. This principle states that
no two fermions ($e.g.$ electrons, protons, $etc$.) can possess identical
quantum numbers (like spin, charge, angular momentum, $etc$.) at the
same space-time point.
\par
The remnant core contraction implies a high density situation for the
material. That is, the constituents of the core are squeezed into
smaller and smaller volumes of space. When the pressure brought on by
the high density exceeds the electrostatic binding energy of the
electrons to the atomic nuclei, the bound electrons will get detached
from the nuclei and form a gas of free electrons. Because the
electrons must obey the Pauli exclusion principle, the net result of
the density squeeze is that the dense electron gas will behave as if
it is a stiff ball of steel, $i.e.$ it cannot be further
compressed after a certain point. In other words, a dense
electron gas will exert a pressure that will resist
gravitational collapse. This pressure is not thermal but
entirely quantum mechanical in origin, and is called degeneracy
pressure. Configurations of the remnant core whose degeneracy
pressure balances the gravitational attraction are called white
dwarfs. At this configuration, the diameter is very small (to
correspond to the requisite high density), about $1\%$ of the
solar diameter, $i.e.$ comparable to the earth's diameter, and is
very hot. From this point in time onwards, the white dwarfs
gradually cool down. Because of the small surface area and the
peculiarity of its composition, the time scale of cooling of a
white dwarf is very large, several billions of years. Typical
densities of white dwarfs are $(10^6 - 10^9)~g~cm^{-3}$.
\par
For stars with initial mass in excess of $6 M_{\odot}$, the core
is sufficiently massive so that when it undergoes gravitational
contraction at the end of helium burning stage, temperatures
become so high that new fusion processes can occur. These
produce heavier atomic nuclei : carbon, oxygen, neon, $etc$. ... ,
upto iron. By the time iron nuclei are produced, there is a huge
build-up of Coulomb repulsive forces because of the presence of
a large number of positively charged protons. Furthermore, the
binding energy per particle is very high for iron nuclei, so
that these nuclei are very stable. As a result, no further
fusion reactions are possible, and the core starts to collapse again.
\par
This time around, the gravitational collapse becomes too strong
(because of higher mass), and will overcome the electron
degeneracy pressure. Two final stages of this collapse are
possible : neutron stars and black holes.
\par
According to detailed calculations, when the density of matter
in the remnant core reaches about $10^{12}~g~cm^{-3}$, the
composition is substantial amount of free neutrons (that is,
unbound to nuclei), in addition to nuclei and the dense ambient
electron gas. These neutrons will not undergo beta decay because
of phase space barrier brought on by the presence of dense
surrounding electrons, whose Fermi momentum gets to be pretty
high ($\geq 20~MeV$). Beyond a density of about $10^{14}~g
{}~cm^{-3}$, the nuclei `dissolve' because of depletion of their
proton concentration due to inverse beta decay (which becomes
energetically favourable at high densities), and the composition
of the core becomes mostly neutrons, with a small admixture of
protons and electrons. If the core is not too massive
(calculations suggest an upper limit of 2-3 solar masses), the
degeneracy pressure and the repulsive force of neutron matter
can balance the gravitational attraction. A stable configuration
is then possible, and it is called a neutron star. Because of
the high densities involved, a neutron star is expected to be
very compact by stellar standards, with radius about (10-15) $km$.
Since degeneracy pressure plays an important role for the
stability of white dwarfs and neutron stars, these are sometimes
referred to as degenerate stars (in which all fusion reactions
have stopped, unlike ordinary stars).
\par
If the mass of the collapsing core is more than 3$M_{\odot}$,
the collapse will go unchecked. There is no physical
mechanism known that can provide enough repulsive forces to halt
the collapse. This will lead to a singularity situation, where
the density will be extremely high with a large surface
gravity such that not even light radiation can emerge from its
surface. This is called a black hole.
\par
Immediately after the discovery of neutron in 1932, it was visualized
by Landau that very compact stable objects, made up entirely of
neutrons could be formed in stellar collapse. Baade and Zwicky \c{1a}
in 1934 suggested that such dense objects could be the remnants formed
in the aftermath of supernova explosions. The first quantitative
theoretical
estimates for the bulk properties of neutron stars, such as mass and
radius, were given by Oppenheimer and Volkoff \c{1}. Although the
existence of neutron stars were predicted in 1932 and the theoretical
estimates of their mass and radius were done in 1939 (Oppenheimer and
Volkoff \c{1}), it was not until the discovery of pulsars \c{2} in 1968, that
the real astrophysical importance of the neutron star idea was
established. It is now generally accepted that pulsars are rotating,
magnetized neutron stars. In order to have a proper understanding of
the structure and the dynamic of pulsars, it is necessary to study
the structure and composition of neutron stars in detail.
Besides, the maximum mass possible for stable neutron stars is
important in another astrophysical situation, namely the
identification of a possible black hole in compact binary systems.
\par
Typical neutron star mass and radius are expected to be about
1$M_{\odot}$ ($M_{\odot} = $ solar mass $= 2\times 10^{33}~gm$) and
10~$km$ respectively. This means that the average density inside a
neutron star is of the order $\rho_{0}$ or higher, where $\rho_{0}=$
nuclear matter density $\simeq 2.8\times 10^{14}~g~cm^{-3}$. Most of
the matter inside a neutron star is expected to be at densities much
above $\rho_{0}$. The behaviour of such matter is not well understood.
{}From heavy-ion collision data at intermediate energies, one now hopes
to derive reliable nuclear equation of state for nuclear matter for
$\rho = (2-3)\rho_{0}$ \c{3}.
The interactions that will be important beyond this
density are not known, although several theoretical models have been
proposed during the last twenty years. An additional difficulty is the
many-body aspect of the problem. A proper way to incorporate the
many-body correlation effect in high density strongly
interacting matter is necessary, but
there exists as yet no consensus on this subject.
\par
There is general acceptance that neutron star interiors can be
divided into four distinct density regimes characterized by
different compositions  \c{4,5} :
\begin{enumerate}
\item The  density  at the surface made up
basically of $^{56}Fe$ nuclei bound in a lattice immersed in a sea of
relativistic nondegenerate electrons, which ranges from
$7.86~g~cm^{-3}$ to a density of $\sim 10^{7}~g~cm^{-3}$.
\item The second density region is between $10^{7}~g~cm^{-3}$ to
$4.3\times 10^{11}~g~cm^{-3}$. The latter corresponds to the
neutron drip point, at which point onwards continuum neutron states
start getting populated.
\item The third density region starts at $4.3\times
10^{11}~g~cm^{-3}$ and it continues all the way upto nuclear
matter density of $\rho_0 = 2.8\times 10^{14}~g~cm^{-3}$. In
this density regime, the
matter consist of neutron--rich nuclei, neutrons (expected to be
superfluid) and
small number of protons and electrons.
\par
The density regions (1), (2) and (3) together define the crust
of the neutron star because the nuclei are expected to be in
lattice structure. The crustal depth from the surface depends on
the equation of state and roughly corresponds to about $(10-15)\%$ of radius.
\item  The fourth density regime corresponds to $\rho\ge\rho_0$.
The central densities can be $\simeq ~ 10~\rho_{0}$.
\end{enumerate}
\par
Oppenheimer and Volkoff \c{1} considered neutron star matter to be
non-interacting neutron matter and obtained a value $0.7M_{\odot}$ as
the maximum gravitational mass and a corresponding radius $\sim
17$~$km$. Today it is believed that neutron star interiors are made
up asymmetric nuclear matter, with perhaps small admixture of
pions and hyperons. Since neutron
star matter contains matter at density higher than the nuclear matter
density, one should take into account, as far as possible, the nuclear
forces at short ranges. In the last twenty years, many attempts have been
made to calculate the equation of state $p~=~p(\rho)$, $p~=~$pressure of high
density matter, taking
into account the nuclear forces at short distances. Most such
realistic calculations give the neutron star maximum
gravitational mass to be between $(1.8 - 2)M_{\odot}$.
\par
Clearly, it is important that interactions among neutrons be included in
any realistic calculation of neutron star structure. The
equation of state where the density $\ge
10~\rho_{0}$ is poorly understood due to uncertainty in nuclear
interactions, and has been a focus of much theoretical research
in recent years. In this thesis, we make a study of properties
of dense nuclear matter using a field theoretical approach,
which has gained increasing importance in the last few years.
\par
This thesis contains two parts. Part one (chapters 2--5)
deals with the nuclear equation of state based on relativistic
mean field theory. In chapter 2, we briefly review the
equations of state in both non-relativistic and relativistic
approaches. In chapter 3, we make a detailed study of the chiral
sigma model and use it to derive an equation of state of
asymmetric nuclear matter. Recently, Datta and Alpar \c{6} showed
that the Vela pulsar glitch data suggest stiffer rather than
softer equation of state for neutron star matter.
The chiral sigma model seems to possess such a desirable
property at high densities. We have extended the chiral
sigma model calculation to finite temperatures
in chapter 4. We also discuss the derivative scalar coupling
model proposed by Zimanyi and Moszkowski \c{7} some years ago.
In chapter 5, we examine in the context of our model (as
discussed in chapter 3), the question of a possible phase
transition from hadronic matter to quark matter at high densities.
We discuss the formation of strangelets of large
mass in quark matter at high densities based on non-relativistic
treatment in chapter 6. The detection of strangelets may be the most
unambiguous way to confirm the formation of quark-gluon plasma in
heavy ion collision experiment. The study of this chapter gives basic
idea on the strangeness contents in the high density matter such as
core of the neutron stars.
\par
In the second part of the thesis, we consider certain
astrophysical applications. These are the modelling of stable
neutron star structure, radial oscillations of quark stars and
cooling rates of neutron stars if these objects
possess ($u,~d,~s$) quark matter in their cores. The structure of
neutron stars, quark stars and quark star oscillations are
discussed in chapter 7. In the chapter 8, we discuss the
limitation of previously suggested neutrino emissivity formula
for the quark matter in the neutron stars. In the last chapter,
we summarize the highlights and the main points of this thesis.
\vfill
\newpage
\setcounter{equation}{0}
\chapter{The equation of state of matter in neutron star
interior : a review of previous work}
\section{Introduction}
\hspace {0.3in} The equation of state $p(\rho)$ ($p$=pressure, $\rho$=total
mass-energy density) above the nuclear matter density ($\rho~>~
3\times 10^{14}~g~cm^{-3}$) plays a crucial role to determine
the equilibrium structure of neutron stars. A considerable amount
of work has already been done in the last two decades on this
subject. In order to set a proper perspective for
the discussion in the following chapters, we give here a brief
review of previous work on equations of state of matter in
neutron star interior. Upto about nuclear
density the equation of state is reasonably well known, but the
central density
of neutron star can be almost an order of magnitude higher. In this
regime, the physics is unclear. Below nuclear density the nuclear
gas is dilute and one can use the perturbative methods, while at much
higher densities we would be in the asymptotically free region of
quantum-chromo-dynamics. Neutron star matter is usually divided
into four general density regions  \c{4,5}:
\par
\begin{enumerate}
\item  Near the surface region, where density is upto $10^{7}~g~cm^{-3}$,
a lattice of bare nuclei (mainly $^{56}Fe$) is immersed in a gas of
relativistic and degenerate electron gas. The equation of state in
this region is influenced by temperature and magnetic fields . But
it hardly matters to the structure of neutron stars.

\item  The next region is the neutron-rich nuclei upto $^{118}Kr$, where the
protons inside the nuclei undergo inverse beta decay. Due to the inverse
beta decay, the electron gas occupies the lowest Fermi energy level,
which allows the system to be in a lower value for the ground state energy.

\item  The third density region begins at about $4.3\times
10^{11}~g~cm^{-3}$ is called the neutron drip point. In this
region, some of the neutrons in the nuclei get detached from the
parent nucleus. These neutrons are unbound and stable.

\item In the final region, (density $>$ $2.8\times
10^{14}~g~cm^{-3}$), the individual nuclei merge into each
other. So, the composition of the matter is expected to be a
fluid of almost uniform
neutron matter together with protons, electrons and possibly muons, pions,
hyperons $etc$. Above this density, there is possibility of a phase
transition from neutron matter to ($u,~d,~s$) quark matter.
\end{enumerate}
\par
Because of large neutrino emissivity, neutron star matter is
cold and degenerate. Therefore, we do not have to worry about
temperature in deriving equation of state.
\par
The first three of the above density regions are reasonably
well-understood. Comprehensive account of the physics of these
regions are given by Canuto \c{4} and Baym and Pethick \c{5}. All
versions of equation
of state of the fourth region (which is above the nuclear matter
densities) differ from each other, due to the inadequate knowledge of
nuclear interactions as well as the lack of proper many--body
techniques, that will be relevant to describe neutron star matter.  In
what follows, we briefly review the two major techniques that have been
used :

\begin{enumerate}
\item  non-relativistic theories,
\item  relativistic field theoretical model.
\end{enumerate}
\section{Non-relativistic approach}
\hspace{0.3in} In this approach one uses two-body potentials
which are fitted to the nucleon-nucleon scattering data, as well
as a three-body term. The form of the potential is chosen whose
parameters are determined using the data of few-body nuclei and
saturation properties of nuclear matter. For a non-relativistic model, the
starting point is the two-body potential. For the many--body method, there
are two approaches, Brueckner-Bethe-Goldstone theory \c{8}
and the variational method \c{9}. Since the major constituents of matter inside
the neutron stars are neutrons, we concentrate our attention on
neutrons and their interactions. The experimental information
regarding nucleon-nucleon scattering data and known properties
of deuteron do not
uniquely determine the nucleon-nucleon potential. Hence, for setting
up the equation of state, it is required to have fit with known
properties of equilibrium nuclear matter at saturation density such
as binding energy per nucleon, compression modulus $etc$. Before we
discuss the recent non-relativistic equation of state given by Wiringa et
al. \c{10} that provides a reasonably good description, we briefly
review the historical equations of state at high
densities in the following sections.
\par
\ni {\Large\it I. Reid Model:}
\par
The Reid model \c{11} is based on phenomenological nucleon-nucleon
potentials, which fits the scattering data very well. It has
been used
extensively in the calculation of neutron star structure (reviews by
Baym and Pethick \c{12}; Canuto \c{13}). The calculation of
equation of state $p(\rho)$ using the lowest order constrained
variational method seems to be accurate enough for the central part of
the Reid potential (Pandharipande and Bethe \c{14}). But
Pandharipande and Wiringa \c{15} calculated the nuclear matter properties
using Reid
potential and found that both the equilibrium density as well as binding
energy were too large. Hence, this model is now considered to be
unrealistic.
\par
\ni {\Large\it II. Bethe-Johnson model:}
\par
The short-range interaction between nucleons is not uniquely
determined by the nucleon-nucleon scattering data. So, Bethe and
Johnson \c{16}  proposed a phenomenological potential model
where they suggested
several different potential models for nucleon-nucleon interaction, assuming
various plausible strengths for short-range repulsion by fitting the
scattering data.
The maximum mass calculations for neutron stars with this equation of
state are given in Malone $et~al.$ \c{17}. At high densities the
constituents of matter are made up of nucleons ($N$) and hyperons
($Y$). However, the
hyperonic interactions are not included in full details.
\par
These two models (I and II) are phenomenological density independent static
potential models. According to these, the nuclear matter energy at high
densities increases linearly with the density.
\newpage
\par
\ni {\Large\it III. Tensor Interaction model:}
\par
The attraction between nucleons which comes from higher order contribution
of the pion exchange tensor interaction was studied by Pandharipande
and Smith \c{18}. Their work is a generalization of
various tensor interaction models proposed earlier by Green and
Haapakoski \c{19}.
These interactions fit only the $s$-wave scattering data, and differ mainly
in the strength of the short-range repulsion for which a
specific form is presumed.
The work by Smith and Pandharipande in \c{20}
suggests that the low energy nucleon-nucleon scattering data can
be explained by
attributing all the attraction between nucleons to tensor
interaction. It has been seen that the calculation with tensor
interaction using lowest order variational and Brueckner methods
(Green and Niskanen \c{21}) satisfies only half of the nuclear
matter binding energy at saturation density. Thus the tensor
interaction model can not
explain the nuclear matter properties in a satisfactory way.
\par
\ni {\Large\it IV. The mean field model:}
\par
Pandharipande and Smith \c{22} assumed a model, called the mean
field model,
which states that the attraction between nucleons is due to the
exchange of an effective scalar meson. The quadrupole moment of the
deuteron as well as the phase shift in the $^{3}P_0,~^{3}P_1$ and
$^{3}P_2$ channels clearly indicate the presence of the one-pion
exchange tensor force between the nucleons.
However, a detailed analysis of the attractive interaction due
to all possible tensor potentials
(Smith and Pandharipande \c{20}) suggest that it is
almost independent
of the spin and isospin of the interacting nucleons, and thus
its contribution in matter could be similar to that due to
coupling of nucleons to a scalar field. The scalar field, treated in
the mean field approximation, was used by Walecka \c{23}, which
we shall discuss later in detail. Following the Walecka model,
the nucleons moving in a mean scalar field are assumed to interact by
a central potential generated by $\omega$, $\rho$ and $\pi$ mesons
exchanges. The potential approximation may be appropriate for
$\omega$ and $\rho$ vector fields, while the central parts of the pion
exchange potentials have a negligibly small effect.
The mean field
approximation is unjustified for the vector field because the
$\omega$-$\rho$ exchange potential have a range $\sim$ 0.2 $fm$,
which is much smaller than the mean interparticle spacing $\sim$
1.2 $fm$. The
short-range corrections induced by the $\omega$-$\rho$ exchange
potentials are treated by variational method using a hypernetted chain
formalism (Pandharipande and Bethe \c{24}). The coupling constant of
$\omega$, $\rho$ and $\sigma$ are calculated from the nuclear matter
binding, symmetry energy and the equilibrium density. The
incompressibility parameter obtained by this theory is $\sim$ 310
$MeV$. The nucleon-nucleon scattering data can not be explained by
an interaction which has been
used in the mean field model. But this model satisfies all the
empirically computed properties of nuclear matter.
\par
The attraction between nucleons in the above two models, $i.e.$, tensor and
mean field decrease with increasing density. However, this is
the  general characteristic of microscopic models, which are
based on the mean field theoretic calculations. The draw back
of these two models is that at small densities the energy is
proportional to the density, whereas at higher densities it tends to
saturate.
\par
\ni {\Large\it V. Friedman and Pandharipande model:}
\par
Friedman and Pandharipande \c{27} gave a model for the equation of
state of dense neutron and nuclear matter, where they used the
variational method as suggested earlier by Pandharipande \c{9}
to calculate the equation of state for a wide range of density. In
their model, they used improved phenomenological nucleon-nucleon
interactions \c{27a,27b}. The two nucleon interaction has short--range and
intermediate range parts, and also possesses the pion--exchange
contribution. There is also a contribution due to a
three--nucleon interaction which has a complicated form (it is
a function of strength parameters, the
interparticle distance and also the alignment angles). The parameters
involved in the three--nucleon interactions are determined by reproducing
the equilibrium density, energy and incompressibility of nuclear matter
based on variational calculations. This model fits well the
nucleon-nucleon scattering cross-section data, the deuteron properties and
the nuclear matter properties.
\par
\ni {\Large\it VI. Wiringa, Fiks and Fabrocini model:}
\par
Wiringa $et~al.$ \c{10}  proposed a model which is most firmly based
on available nuclear data. This model improves on the earlier work by
Friedman and Pandharipande \c{27}. In this model the two-nucleon
potential is taken to be the Argonne v14 (AV14) or Urbana (UV14)
potential. Both have the identical structure, but differ in the strength of
the short-range tensor force. They are called v14 models because they
are the sum of v14 operator components (like
$\sigma_{i}.\sigma_{j},~\pi_{i}. \pi_{j}$, $etc$.). Each of these
components has three radial pieces which includes the long-range
one-pion exchange, an intermediate-range part that comes from the
two-pion exchange processes, and a short-range part, coming either from the
exchange of heavier mesons or overlaps of composite quark systems.
All the free parameters are fitted to nucleon-nucleon scattering
data and deuteron properties.
\par
For the three-nucleon interaction, the Urbana VII potential is used,
which has a two-pion exchange part and intermediate-range repulsive
contribution. Calculations have been carried out for the Urbana
$v_{14}$ plus three-nucleon interaction (TNI) model of Lagaris and
Pandharipande \c{28}.
\par
The many-body calculations are based on the variational principle
where one uses the technique called the Fermi hypernetted
chain-single-operator chain (FHNC $-$ SOC) integral equations \c{27a,29}.
\par
The authors obtained the nuclear matter saturation properties such as
binding energy, saturation density and incompressibility parameter
value of nuclear matter at saturation density are --16.6 $MeV$,
$0.157~fm^{-3}$ and 261.0 $MeV$ respectively for UV14 plus TNI model.
This is a reasonable improvement over all the other non-relativistic
approaches.
\par
In their study, one notices that the sound velocity $s$ in the medium
given by the equation of state (in parametric form) for
beta-stable ($n,~p,~e$ $i.e.$, no hyperons) matter based on
non-relativistic approach violates causality
above $\rho = 1fm^{-3}$. This is an undesirable feature of this
method at high densities. However, they predicted the maximum neutron
star mass to be $2.2M_{\odot}$ and for 1.4$M_{\odot}$ neutron star,
the central density turns out to be substantially below
$1fm^{-3}$ which is quite realistic. It may be noted that the modern
potentials supplemented with reasonable three-body interactions yield
very similar models of neutron star structure parameters. For the
$1.4M_{\odot}$ models, the radius is $\simeq 10.4 - 11.2~ km$, and
the central density is about $6\rho_{o}$.
\section{The relativistic approach}
\hspace{0.3in} The shape of baryonic potential is not known at
very small
interparticle separations ($\le 0.5 fm$). Also, it is not clear that
the potential description will continue to remain valid at such
short ranges. Moreover, in the neutron star interiors, the Fermi
momentum of the
degenerate neutron is large. Therefore, the non-relativistic approach
may not be adequate. In the recent times, the relativistic approach
has drawn considerable attention.
\par
In the relativistic approach, one usually starts from a
local, renormalizable field theory with baryon and explicit meson degrees
of freedom. These models have the advantage of being relativistic,
however, one drawback is that one does not know how to relate it to
nucleon-nucleon scattering data.
The theory
is chosen to be renormalizable in order to fix the
coupling constants and the mass parameters by
empirical properties of nuclear matter at saturation ( binding,
density, compression modulus, effective mass and symmetry
energy). As a starting point, one chooses the {\it mean field
approximation} (MFA) which should be reasonably good at very
high densities (a few times nuclear) \c{30}. In the second step, one
includes one-loop vacuum fluctuations which leads to what is
called the {\it relativistic Hartree approximation} (RHA) \c{31}.
This approach is currently used as a reasonable way of
parametrizing the equation of state.
\par
The main features of this approach  can already be seen in a simple
model. Assuming that a neutral scalar meson field ($\phi$) and a neutral
vector field ($V_{\mu}$) couple to the baryon current by interaction
terms of the form

\begin{equation}
g_{s}\bar\psi\psi\phi ~~ and ~~ g_{v}\bar\psi\gamma^{\mu}\psi
V_{\mu},\nonumber\\
\end{equation}

\ni the Lagrangian can be written as ($\hbar~=~1~=~c$)

\begin{eqnarray}
{\cal L} =
\bar\psi[\gamma^{\mu}(i\partial_{\mu}-g_{v}V_{\mu})-(M-g_{s}\phi)]\psi
+ {1\over
2}(\partial_{\mu}\phi\partial^{\mu}\phi-m_{s}^{2}\phi^{2}) \nonumber\\
-{1\over 4} F_{\mu\nu}F^{\mu\nu}+{1\over 2}
m_{v}^{2}V_{\mu}V^{\mu};
\end{eqnarray}

\ni where
\begin{equation}
F_{\mu\nu} = \partial_{\mu}V_{\nu}-\partial_{\nu}V_{\mu}.
\end{equation}

\ni From Euler-Lagrange equation

\begin{equation}
{\partial\over {\partial x^{\mu}}}\big[{\partial {\cal
L}\over{\partial(\partial q_i/\partial x^{\mu})}}\big] - {\partial
{\cal L}\over {\partial q_i}}=0
\end{equation}

\ni where $q_i$ is one of the generalized coordinates, one yields the
field equations

\begin{eqnarray}
(\partial_{\mu}\partial^{\mu}+{m_{s}}^{2})\phi=g_s\bar\psi\psi,
\nonumber \\
\partial_{\mu}F^{\mu\nu}+{m_{v}}^{2}V^{\nu}=g_v\bar\psi\gamma^{\nu}\psi,
\nonumber \\
\big[\gamma^{\mu}(i\partial_{\mu}-g_v V_{\mu})-(M-g_s\phi)\big]\psi=0.
\end{eqnarray}

\ni The MFA consists in replacing the meson-field operators by
their expectation values :

\begin{equation}
<\phi> \equiv \phi_{o},~~ < V_{\mu} > \equiv \delta_{\mu o}V_{o},
\end{equation}

\ni if we consider a static  uniform system. Then, the meson field
equations become

\begin{equation}
\phi_{o}={g_{s}\over m_{s}^{2}}<\bar\psi\psi>\equiv{g_{s}\over
m_{s}^{2}}\rho_{s},
\end{equation}

\begin{equation}
V_{o}={g_{v}\over m_{v}^{2}}<\psi^{\dagger}\psi>\equiv{g_{v}\over
m_{v}^{2}}\rho_{B}.
\end{equation}

\ni In this approximation, the nucleon field operator satisfies
a linear equation :

\begin{eqnarray}
[-i\gamma^{\mu}\partial_{\mu}+g_{v}\gamma^{o}V_{o}+M^{*}]\psi=0,
\end{eqnarray}
\ni where $M^{*}~=~M-g_{s}\phi_{o}$ is the effective mass of the
nucleon and the Lagrangian density takes the form

\begin{equation}
{\cal
L_{MFT}}=\bar\psi[i\gamma_{\mu}\partial^{\mu}-g_{v}\gamma^{o}V_{o}-M^{*}]\psi
-{1\over 2}m_{s}^{2}\phi_{o}^{2}+{1\over 2}m_{v}^{2}V_{o}^{2}.
\end{equation}

\ni The quantization here in this case is straightforward. In the
mean field approximation, the conserved energy--momentum tensor is

\begin{equation}
T_{\mu\nu} = -g_{\mu\nu} {\cal L_{MFT}}+{\partial \psi\over{\partial
x^{\nu}}}{\partial {\cal L_{MFT}}\over{\partial (\partial \psi/\partial
x_{\mu})}}.
\end{equation}

\ni Assuming that the bulk neutron star matter can be treated as a
uniform and perfect fluid, we can write the energy
density of the system as

\begin{equation}
\epsilon=<T_{oo}>
\end{equation}

\ni and the pressure as

\begin{equation}
p = {1\over 3} < T_{ii}>.
\end{equation}

\ni The ground state of nuclear matter is characterised by Fermi momentum
$k_{F}$ which is related to the baryon density $\rho_{B}$ as :

\begin{equation}
\rho_{B}={\gamma\over{(2\pi)^{3}}}\int_{0}^{k_F}
d^3k={\gamma\over{6\pi^2}}k_F^{3}.
\end{equation}
\ni Here $\gamma=4$ for symmetric nucleon matter, $\gamma=2$ for neutron
matter. The mean field $\phi_o$(or $M^{*}$) is determined
self-consistently as

\begin{equation}
M^{*}=M-g_{s}\phi_{o}=M-{g_{s}^{2}\over
m_{s}^{2}}\rho_{s}=M-{g_{s}^{2}\over m_{s}^{2}}
{\gamma\over{(2\pi)^{3}}}\int_{0}^{k_F} d^3k
{M^{*}\over{E^{*}(k)}},
\end{equation}
\ni where $E^{*}(k)=(k^{2}+{M^{*}}^{2})^{1/2}$. This leads to a
transcendental equation for a given $k_{F}(\rho_{B})$.

\ni It turns out that only the ratio $g_{s}^{2}/m_{s}^{2}$ and
$g_{v}^{2}/m_{v}^{2}$  enter in the equation of state. These parameters
are fixed such that the nuclear matter will give rise to the correct binding
energy and saturation density :

\begin{equation}
({{E-BM}\over B})_{o}=-15.75~ $MeV$ ,
\end{equation}

\begin{equation}
k_{F}^{o}=1.42~ fm^{-1}~(\gamma=4).
\end{equation}

\ni Thus

\begin{equation}
C_{s}^{2}\equiv g_{s}^{2}({M^{2}\over m_{s}^{2}})=267.1,
\end{equation}

\begin{equation}
C_{v}^{2}\equiv g_{v}^{2}({M^{2}\over m_{v}^{2}})=195.9.
\end{equation}
\ni With these values one can compute the equation of state for neutron
matter. In this simple model there is
a first order phase transition (similar to the liquid-gas transition
in the van der Waals' equation of state). At asymptotically
high densities, the
velocity of sound approaches the velocity of light.
\par
One can make the model more realistic by including for instance, the
interaction of
$\rho$- mesons (charged vector mesons). In such model (QHD-II of
Serot and Walecka \c{31}) the $\rho$ meson stiffens the equation of state at
relatively low density and causes the gas-liquid phase transition to
disappear.
\par
The above field theoretical treatment is now-a-days referred to
as the Walecka model \c{23}. In 1974, Walecka first proposed this
mean field model, where he
chose the coupling constants in such a way that it fitted the nuclear
matter binding energy and saturation density. In this model,
the value of nuclear matter incompressibility at saturation is
quite high. The isospin triplet vector
meson $\rho$, was not included, but is of relevance in neutron
star interior ($n,~p,~e$) matter. The extension of this model
will be discussed in the next section.
\par
\ni {\Large\it I. Glendenning Model:}
\par
Glendenning \c{30}  presented a relativistic field theoretical model
for densities near as well as above the nuclear matter density. This
includes isospin-asymmetric baryon matter. In this model was
included the nucleons, the mesons : $\omega$, $\rho$ and
$\pi$, $e^{-}$ and $\mu^{-}$ particles along with the self-interaction
of the scalar meson field $\sigma$. The self-interaction of
$\sigma$-field was chosen to be of the form :

\begin{equation}
U(\sigma) = ({b\over 3} m_n + {c\over 4}
g_{\sigma}\sigma)(g_{\sigma}\sigma)^{3}
\end{equation}

\ni where $m_n$ is the nucleon rest mass and $b$, $c$ are parameters. The
coupling constants and the parameters $b$, $c$ were determined by
satisfying the following bulk properties of nuclear matter $i.e.$,

saturation particle density =  0.145 $fm^{-3}$

saturation binding energy = --15.96 $MeV/nucleon$

asymmetry energy  = 37 $MeV$

and the incompressibility = 280 $MeV$
\par
The author chose two different sets of parameters, one of which
gives a soft and other a stiff equation of state at high densities,
for fits to nuclear physics.
\par
In 1985, Glendenning \c{32} extended  the above formalism to derive an
equation of state of hyperonic matter. He used the same constraints
as earlier to fix the various parameters. Here, he included the $K$
and $K^{*}$ meson exchanges and also the self-interaction form as earlier.
The coupling strengths of the isobar $\Delta$ and the hyperons
($\Lambda$, $\Sigma^{\pm,0}$, $\Xi^{\pm,0}$) to the mesons are
taken from the work of Moszkowski \c{33}. This work  indicates that
cores of the heavier neutron stars are dominated by hyperons and the
total hyperon population for such stars are $15\%-20\%$, depending on
whether pions condense or not. The advantage of this model is that one
can make the equation of state soft or stiff depending on the input
parameters, namely the incompressibility value. In this case the following
four separate hyperonic equation of state matter are considered.

Case 1 : $n$, $p$, hyperons, $\Delta$, $e^{-}$, $\mu^{-}$, $\pi^{-}$;

Case 2 : $n$, $p$, hyperons, $\Delta$, $e^{-}$, $\mu^{-}$;

Case 3 : like case 2 but with universal coupling of hyperons;

Case 4 : like case 3 but no $\rho$- exchange ($g_{\sigma}=0$).
\par
We describe now the results of a more recent analysis \c{34} for the
following ($\sigma,~\omega,~\rho$) model, whose Lagrangian is given as

\begin{eqnarray}
{\cal L }=  \sum_{B}\bar\psi_{B}(i\gamma^{\mu}\partial_{\mu}-m_{B}-g_{\sigma
B}\sigma - g_{\omega B}\gamma^{\mu}\omega_{\mu} - {1\over 2}g_{\rho
B}\gamma^{\mu} \tau_{3}\rho_{\mu}^{(3)})\psi_{B} \nonumber \\
 + {1\over 2}(\gamma_{\mu}\sigma\gamma^{\mu}\sigma -
m_{\sigma}^{2}\sigma^{2})-{1\over
4}\omega_{\mu\nu}\omega^{\mu\nu}+{1\over 2}
m_{\omega}^{2}\omega_{\mu}\omega^{\mu} \nonumber \\
 - {1\over
4}\vec\rho_{\mu\nu}\vec\rho^{\mu\nu}+{1\over
2}m_{\rho}^{2}\vec\rho_{\mu}.\vec\rho^{\mu} -
{1\over3}bm_{n}(g_{\sigma}\sigma)^{3} - {1\over
4}c(g_{\sigma}\sigma)^{4}.
\end{eqnarray}
Here the sum runs over the baryons $N,~\Lambda,~\Sigma,~\Xi,~\Delta,~ etc$.
\par
For uniform matter, the theory depends only on the ratios
$g_{\sigma}/m_{\sigma},~g_{\omega}/m_{\omega},~g_{\rho}/m_{\rho}$ and
the self-interaction coefficients $b,~c$. These five quantities are determined
empirically, in  both the MFA and the RHA ( Table 1 in Ref.
\c{34}).
{}From this calculation one notices that the equation of state for
stable neutron star matter will be softer than the pure
neutron matter. This
softening is the result of the replacement of energetic neutrons
by hyperons at rest. It is most
remarkable that the maximum allowed neutron star mass depends
very sensitively
on this coupling. This point has been
further investigated in Ref. \c{35}, with the conclusion that even if
we have a perfect knowledge of the nuclear equation of state upto about
$2\rho_{o}$, there is still a large uncertainty in the maximum
neutron star mass
because of our lack of knowledge of the hyperon interactions. This is
one of the most relevant ``interface" of astrophysics and nuclear physics.
However, further laboratory experiments are needed to reduce the large
uncertainties.
\par
\ni {\Large\it II. Alonso and Cabanell model:}
\par
Alonso and Cabanell \c{36} gave a model in which the nucleons
interact via scalar
($\sigma$) mesons, pions ($\pi$), and vector ($\omega$ and $\rho$)
mesons. The model is solved in the renormalized Hartree
approximation. The authors give two sets of equation of state
(I, II).
\par
The equation of state I is derived by fitting the properties of
symmetric nuclear matter at nuclear density in a satisfactory
way. This is in good agreement with the equation of state
obtained by Baym, Bethe and Pethick \c{37} in the region above the
neutron drip for neutron
matter. The value of the nuclear incompressibility obtained in this
model is too high (460 $MeV$).
This large incompressibility is perhaps due to the
absence of $\sigma$- self-interaction term in the Lagrangian.
\par
The equation of state II is derived using the chirally invariant
$\sigma$- model Lagrangian, coupled to $\omega$ and $\rho$ mesons and
having an explicit symmetry breaking term. This equation of state
fits accurately all the known properties of symmetry energy of
nuclear matter
at nuclear density. The value of the nuclear incompressibility
at saturation density comes out to be 225 $MeV$.
\par
For these two equations of state, the total number of adjustable
parameters are rather large (12 for set I and 15 for set II). A
peculiar feature of these models is that both (I, II) show a dip in equation
of state. These dips do not represent any phase transition for the
system.
\par
\ni {\Large\it III. Chiral sigma model (Glendenning):}
\par
Chiral symmetry is a good hadron symmetry, ranking only below the
isotopic spin symmetry \c{38}. For this reason, it is
desirable to have chiral symmetry in any theory of dense
hadronic matter. At
the same time, the theory should be capable of describing the bulk
properties of nuclear matter. But a theory that satisfies both the
conditions is not available.
\par
In 1986, Glendenning \c{39} derived the equation of state based on the mean
field approximation. Here, the equation of motion for the mean fields
is also derived in the general case for the chiral sigma model
supplemented by a
gauge massless vector meson in interaction with the other hadrons,
including baryon resonances. Here, omega meson is considered
with dynamical masses. In this theory, the gauge field $\omega_{\mu}$
of massless vector meson is introduced into the chiral sigma model
through the covariant derivatives. Moreover, the linear term of
$\sigma$ field in it is the symmetry breaking term, by which the pion
acquires a finite mass. In addition to the $\omega$-mesons, the theory
has scalar meson and the pseudoscalar pion.
This theory has two parameters, which are determined
from the saturation density and binding energy per nucleon in normal
symmetric matter. The value of the incompressibility is rather very
large $i.e.$, $K = 650$ $MeV$.
\par
In a subsequent paper, Glendenning \c{40} extended the chiral sigma model
based on the mean field approximation where the $\omega$-meson does not
have a dynamically generated mass, even if one considers the vacuum
renormalization correction to the theory. In this model, he considered
only the normal non-pion condensed state of matter and
included hyperons in beta equilibrium with nucleons and leptons. He fitted
the parameters of the theory to obtain two equation of state corresponding
to two compression moduli, a ``stiff" ($K=300~ MeV$) and a ``soft" ($K=200
{}~MeV$), by reproducing correct nuclear matter properties.
\newpage
\par
\ni {\Large\it IV. Chiral sigma model (Prakash and Ainsworth):}
\par
Prakash and Ainsworth \c{41} proposed an equation of state
based on  the chiral
sigma model. They examined the role of the many-body effects provided by
the chiral sigma model in the equation of state of symmetric nuclear
matter and neutron rich matter. They include the  $\sigma$- meson
one-loop contributions, but the isoscalar vector field is not
generated dynamically, so that its role is reduced to an empirical
one. A set of equation of state is constructed, each of which fits
the empirical saturation density, the binding energy and the
symmetry energy of
nuclear matter. They obtained a value of compression modulus at
saturation nuclear matter density for symmetric matter that is different
from the experimental expected value. So, they allowed a
variation in the values of coupling constants arbitrarily so
that the theory will give rise to the desired
value of compression modulus. The vector field plays no role in
determining the value of the effective mass of the nucleon in such an
approach.
\par
\ni {\Large\it V. Baron, Cooperstein and Kahana model:}
\par
Baron $et~al.$ \c{42} gave a phenomenological model for neutron
matter equation of state.
The form of the nuclear partial
pressure is

\begin{equation}
P~ =~ {{K_0\rho_0}\over{9 \gamma}}[ u^{\gamma}-1],
\end{equation}

\ni where the baryon density compression factor $u=\rho/\rho_0$, measured
with respect to the empirical symmetric nuclear matter
saturation density, $\rho_0 = 2.8\times 10^{14}~g ~cm^{-3} =
0.16~fm^{-3}$. Here, $K_0$ is the nuclear incompressibility at
saturation, $i.e.$ $u=1$, while $\gamma$ is the extremely high-density
adiabatic index.
\par
Later, Cooperstein \c{43} modified the phenomenological equation of
state to the following form

\begin{equation}
P~ =~ {{K_N\rho_0}\over{9 \gamma}}u^{\gamma},
\end{equation}

\begin{equation}
\varepsilon~ =~ \rho_0(m_n+m_1)u~ +~ {P\over{\gamma -1}},
\end{equation}

\ni where the integration constant $m_1$ is given by

\begin{equation}
m_1 ~ =~ E_{nm}+E_{sym}-{K_N\over{9\gamma(\gamma -1)}},
\end{equation}

\ni with the empirical value $E_{nm}=-16 MeV$ and for
definiteness $E_{sym}=36$
$MeV$. Here the factor $u^{\gamma}$ is replaced by $u^{\gamma}-1$ for
nuclear matter saturation. The neutron matter compression modulus is
denoted by $K_N$ and $m_n$ is the nucleon mass. The value of $K_N$ is
not interpreted as the correct
saturation density but is represented as a high density neutron
matter parameter. This is connected with $K_0$ through the density
dependence of the symmetry energy. One can find different equations of
state by varying the values of $K_N$ and $\gamma$ in this theory.
For particular values of $K_N$ and $\gamma$, these phenomenological
equations of state fit well the non-relativistic equation of
states, such as that of the Bethe and Johnson \c{16} or Friedman and
Pandharipande \c{27}.

\ni {\Large\it VI. Rosenhauer $et~al.$ model:}
\par
Rosenhauer $et~al.$ \c{44}  refer two commonly used parameterizations for
the hadronic equation of state, namely that of Sierk and Nix \c{45}

\begin{equation}
\varepsilon_{SN}(n)~=~ {{2 K}\over 9}(\sqrt{n/n_0}-1)^{2}
\end{equation}

\ni and the quadratic form \c{46} is

\begin{equation}
\varepsilon_{Q}(n)~ =~ {K\over 18}(n-n_0)^{2}/{n_{0}}^{2}.
\end{equation}
\ni The symbol $\varepsilon_{SN}(n)$ and $\varepsilon_{Q}(n)$
represent energy density of the two equations of state, namely, Sierk
and Nix as well as quadratic form respectively.
Here, $n_0$ is the normal nuclear matter density and $K$, the
compression modulus characterizing the properties of nuclear matter
at densities $n>n_0$. In these two parametric equations,
$\varepsilon_{comp}(n)$ ($i.e.$, $\varepsilon_{SN}(n)$ or
$\varepsilon_{Q}(n)$), the large $K$ value indicates the stronger
repulsive nature of nucleon-nucleon interaction.

\ni The total energy density is then expressed as

\begin{equation}
\varepsilon (n)~ =~ n[\varepsilon_{comp}(n)+W_0+m_n+W_{syms}],
\end{equation}

\ni where the binding energy per nucleon at normal nuclear matter
density, ($n_0=0.145~fm^{-3}$) is $W_0=-16~ MeV$, $m_n$ being the rest
mass of the neutron and $W_{sym}=32~ MeV$, the symmetry energy of
neutron matter estimated from the liquid drop model. The symmetry
energy $W_{sym}$ and the binding energy $W_0$ determine the
properties of matter at saturation within this phenomenological
approach. However, $\varepsilon_{comp}(n)$ incorporates all density
dependent effects.
\par
\ni {\Large\it VII. Zimanyi and Moszkowski model:}
\par
Zimanyi and Moszkowski \c{7} proposed a model similar to the
Walecka model but using the derivative scalar coupling (DSC) for
the scalar field. The interesting feature of this model
is that the equation of motion of the scalar field becomes non-linear
without introducing any extra parameter.
This model gives rise to a reasonable value
of the effective nucleon mass and satisfactory value for the incompressibility
of nuclear matter at saturation density.
\par
Glendenning $et~al.$ \c{47} extended the above model to hyperonic
matter and applied to compute a number of neutron star properties. In
place of the purely derivative coupling of the scalar field to the
baryons and vector meson fields of the above (DSC) model, they here
coupled it by both Yukawa point and derivative coupling to
baryons and both vector fields. This improves the value of the
compression modulus
and effective nucleon mass at saturation density compared to that
obtained in earlier calculation. Also, they include the $\rho$- meson to
account for the asymmetry effect.
\vfill
\newpage
\setcounter{equation}{0}
\chapter{Chiral sigma model}
\section{Introduction}
\hspace {0.3in} The chiral symmetry is a good hadron
symmetry, which ranks
only below the isotopic spin symmetry \c{38}. Due to this
reason, it is expected that any theory of dense matter should possess
it. If one considers the chiral symmetry in dense matter theory,
it should be capable of describing the bulk properties of
nuclear matter such as binding energy per nucleon, saturation
density, compression modulus and symmetry energy. So far,
there has not been any theory possessing chiral symmetry and
describing all the nuclear matter properties. In recent years, the
importance of the three-body forces in the equation of state at high
densities has been emphasized by several authors \c{49,50}. This gives
theoretical impetus to study the chiral sigma model,
because the non-linear terms in the chiral sigma Lagrangian can give
rise to the three-body forces.
\par
A chiral Lagrangian using the scalar field (the so-called sigma model) was
originally introduced by Gell-Mann and Levy \c{51} as an example to
illustrate chiral symmetry and partial conservation of axial current. The
importance of chiral symmetry in the study of nuclear matter properties
was emphasized by Lee and Wick \c{52}. The non-linear terms of the chiral
Lagrangian can provide the three-body forces, important at high densities
(Jackson, Rho and Krotscheck \c{49}; Ainsworth $et~al.$ \c{50}), and can be
relevant in applications to neutron star structure and supernova collapse
dynamics.
\par
The usual theory of pions leads to a theory of nuclear matter that does not
possess the empirically desirable saturation property.
For this reason, the isoscalar vector field is introduced in the
theory via the Higgs mechanism. This way it becomes possible to have a
saturating nuclear matter equation of state (Boguta \c{53}). With
the availability of experimental estimate for the
incompressibility parameter of nuclear
matter (denoted by $K$), there have been attempts to reproduce the desirable
value of $K$ (about 200 $-$ 300 $MeV$) using the sigma model. In the `standard'
sigma model, the value of $K$ turns out to be quite large, several times the
above-mentioned value for plausible values of the coupling constants
involved, and can be reduced only by introducing in the theory, terms due
to the scalar field self-interactions and/or vacuum fluctuations with
adjustable coefficients.
\par
There have been several earlier papers that have employed the chiral
sigma model to obtain the equation of state of high density matter.
Glendenning \c{39} derives the chiral sigma model equation of state with
normal nuclear matter saturation and then applies to neutron star
structure calculations. He then studies the finite temperature
behaviour of this model. A liquid-gas phase transition occurs in
this model below $T\sim 23 ~MeV$. The equation of motion is
derived in the mean field approximation, supplemented by a gauge
massless vector meson in
interaction with the other hadrons, including baryon resonances. The
gauge field of a massless vector mesons is introduced through the
covariant derivatives into the chiral sigma model. The linear term of
sigma field in the Lagrangian is the symmetry breaking term by which
the pion acquires a finite mass. There are scalar meson and
pseudoscalar pion in the theory, in addition to $\omega$- meson, which
contribute a repulsive force and carry spin and isospin. Also, he
considers this model with dynamical masses but not with the $\rho$- meson
and its isospin symmetry influences. The required parameters in this
theory are calculated from the saturation density and the binding energy
per nucleon in normal symmetric matter. The value of the
compression modulus obtained from this model is too large (650 $MeV$).
\par
In another paper, Glendenning \c{40} extends the chiral sigma model to normal
nuclear matter and neutron star matter by considering the vacuum
renormalisation corrections. The equation of state is derived by
including $\rho$- mesons and hyperons in equilibrium with nucleons and
electrons and muons. In this calculation the $\omega$- meson does not have a
dynamically generated mass. He has considered two sets of coupling
constants corresponding to two compression moduli, a `stiff' ($K=300
{}~MeV$) and `soft' ($K=200~MeV$) equation of states for nuclear matter,
that yield the empirical saturation density, the binding as well
as the symmetry energy. In both cases, the hyperons influence on
the equation of state substantially.
\par
An equation of state based on the chiral sigma model is also
considered by Prakash and Ainsworth \c{41}. They examine the role of the
many-body effects provided by the chiral sigma model in the equation
of state of symmetric nuclear matter and neutron-rich matter.
Previously, Matsui and Serot \c{54} considered the role of only the
nucleon vacuum fluctuation terms at the one loop level. But here,
these authors consider both the nucleon as well as the $\sigma$ meson vacuum
fluctuation terms at one loop level in the chiral sigma model. As
a result, the meson loop shifts the saturation density and thus
incompressibility increases. In this theory, the isoscalar vector field
is not generated dynamically. Its role is an empirical
one. They have constructed a family of equation of states, each of
which fits the empirical saturation density, the binding energy and
the symmetry energy of nuclear matter. They allow variation of the
coupling constants arbitrarily in their calculations by fitting the other
parameters from nuclear saturation properties, so that it is
possible to obtain any desired value of the nuclear matter
compression modulus. Finally, they calculate neutron star
structure using the neutron matter equation of state for $p$, $n$ and
$e$ systems.
\par
We have developed the $SU(2)$ $\times$ $SU(2)$ chiral sigma model \c{54a}
to describe
nuclear matter and neutron star matter. We have adopted an approach
in which the isoscalar vector field is generated dynamically.
Inclusion of such a field is necessary to ensure the saturation property
of nuclear matter. The effective mass of the nucleon thus acquires a density
dependence on both the scalar as well as the vector fields, and must
be obtained self-consistently. We do this using the mean-field
theory where all the meson fields are replaced by their uniform,
expectation values. To describe the nuclear matter we have two
parameters in the theory : (i) the ratio of the coupling
constant to the mass of the scalar and (ii) to the isoscalar
vector fields. This procedure also gives a relatively high value for $K$ at
the saturation density. Although this is an undesirable feature as far as
nuclear matter at saturation density is concerned, it need not be viewed as a
crucial shortcoming for our purpose here in view of the fact that a fit to
$K$ at saturation does not tell us what the slope of the equation of
state should be at densities $\geq 4n_s$ ($n_s =~$saturation
density) (Prakash and Ainsworth \c{41}; Horowitz and Serot \c{55};
Stock \c{56}; Baym \c{57}; Ellis, Kapusta and Olive \c{58}). For
neutron star structure, in which we are most interested as an
application of our equation of state, this high density regime plays the most
important part. The important feature of our work is that the
$\omega$- meson mass and coupling constant are not treated
empirically, but in a fully self-consistent manner. To describe
neutron star matter, we
include the coupling to the isovector $\rho$- meson, the
coupling strength being determined by requiring a fit to the empirical
value of the symmetry energy.
\section{The Model}
\hspace{0.3in} The Lagrangian for an $SU(2)$ $\times$ $SU(2)$ chiral
sigma model that
includes (dynamically) an isoscalar vector field ($\omega_{\mu}$) is (we
choose $\hbar = 1 = c$) :

\begin{eqnarray}
{\cal L} =  \frac{1}{2}\big(\partial_{\mu} \overrightarrow{\pi} .
\partial ^{\mu} \overrightarrow{\pi} + \partial_{\mu} \sigma
\partial^{\mu} \sigma\big) - \frac{\lambda}{4}\big(\overrightarrow{\pi} .
\overrightarrow{\pi} +\sigma^{2} - x^2_o\big)^2\nonumber\\
 - \frac{1}{4} F_{\mu\nu} F_{\mu\nu} + \frac{1}{2}{g_{\omega}}^{2}
\big(\sigma^2 + \overrightarrow{\pi}^2\big) \omega_{\mu}
\omega^{\mu} \nonumber\\
 + g_{\sigma} \bar{\psi} \big(\sigma + i\gamma_5 \overrightarrow{\tau}
. \overrightarrow{\pi}\big) \psi + \bar\psi \big(
i\gamma_{\mu}\partial^{\mu} - g_{\omega}\gamma_{\mu}
\omega^{\mu}\big) \psi,
\end{eqnarray}

\ni where $F_{\mu\nu} \equiv \partial_{\mu} \omega_{\nu} - \partial_{\nu}
\omega_{\mu},~ \psi$ is the nucleon isospin doublet,
$\overrightarrow{\pi}$ is the pseudoscalar pion field and $\sigma$ is the
scalar field. The vector field $\omega_{\mu}$ couples to the conserved
baryonic current $j_{\mu} = \bar{\psi} \gamma_{\mu} \psi$. The expectation
value $<j_o>$ is identifiable as the nucleon number density, which we
denote by $\rho_B$.
\par
The interactions of the scalar and the pseudoscalar mesons with the vector
boson generates a mass for the latter spontaneously by the Higgs
mechanism. The masses for the nucleon, the scalar meson and the vector
meson are respectively given by

\begin{eqnarray}
M = g_{\sigma} x_o ;\nonumber\\
m_{\sigma} = \sqrt{2\lambda} x_o ;\nonumber\\
m_{\omega} = g_{\omega} x_o,
\end{eqnarray}
\ni where $x_o$ is the vacuum expectation value of the sigma field,
$\lambda~=~({m_{\sigma}}^{2}-{m_{\pi}}^{2})/(2 {f_{\pi}}^{2})$
($m_{\pi}$ = pion mass and $f_{\pi}$ is the pion decay coupling
constant) and $g_{\omega}$ and $g_{\sigma}$ are the coupling
constants for the vector and scalar fields respectively.
\par
To derive the thermodynamic quantities of the system of degenerate
nucleons, characterized by the nucleon number density ($\rho_B$) or
equivalently the Fermi momentum $k_F=(6\pi^2 \rho_B/\gamma)^{1/3} (\gamma$
= nucleon spin degeneracy factor), we need to know the dependence of the
meson fields on $\rho_B$. For this, we resort to the mean--field
approximation. This approach has been extensively used to obtain field
theoretical equation of state models for high density matter.
In this approximation,
expected to be valid for degenerate matter at high densities, the mesonic
fields are assumed to be uniform ($i.e.$, space-time independent with no
quantum fluctuations). For the isoscalar vector field, then

\begin{equation}
\omega_{\mu} = \omega_o \delta_{\mu}^o,
\end{equation}

\noindent where $\omega_o$ is space-time independent but depends
on $\rho_B$ and $\delta_{\mu}^o$ is the Kronecker delta. The
equation of motion for the vector field specifies $\omega_o$ :

\begin{eqnarray}
\omega_o = \frac{\rho_B}{g_{\omega}x^2} ,\nonumber\\
x = (<\sigma^2 + \overrightarrow{\pi}^2>)^{1/2}.
\end{eqnarray}

\ni The equation of motion for $\sigma$ is written for convenience in terms
of $y \equiv x/x_o$, and is of the form

\begin{equation}
y(1-y^2) + \frac{c_{\sigma}c_{\omega}\gamma^2k_F^6}{18\pi^4M^2y^3} -
\frac{c_{\sigma}y\gamma}{\pi^2} \int^{k_F}_o \frac{dk k^2}{\big(\overrightarrow
k^2+M^{\star 2}\big)^{1/2}} = 0,
\end{equation}

\noindent where $M^{\star} \equiv yM$ is the effective mass of the
nucleon and

\begin{equation}
c_\sigma \equiv  g_{\sigma}^2/m_{\sigma}^2; \quad\quad\quad
  c_{\omega} \equiv g_{\omega}^2/m_{\omega}^2.
\end{equation}

\noindent  We consider here the normal state of high density matter in which
there is no pion condensation.
\par
The diagonal components of the conserved total stress
tensor corresponding
to the Lagrangian (3.1) together with the equation of motion for the fermion
field (and a mean field approximation for the meson fields) provide the
following identification for the total energy density ($\epsilon$) and
pressure ($P$) of the many-nucleon system (assumed to be a perfect fluid) :

\begin{eqnarray}
\varepsilon = \frac{M^2(1-y^2)^2}{8c_{\sigma}} +
\frac{\gamma^2 c_{\omega}k_F^6}{72\pi^4y^2} + \frac{\gamma}{2\pi^2}
\int_o^{k_F} dk k^2\big(\overrightarrow{k}^2 + M^{\star 2}\big)^{1/2}
\nonumber\\
P = - \frac{M^2(1-y^2)^2}{8c_{\sigma}} +
\frac{\gamma^2 c_{\omega}k_F^6}{72\pi^4y^2} + \frac{\gamma }{6\pi^2}
\int_o^{k_F} \frac{dk k^4}{\big(\overrightarrow{k}^2 + M^{\star 2}
\big)^{1/2}}.
\end{eqnarray}

\ni The energy per nucleon is

\begin{equation}
E = \frac{3\pi^2M^2(1-y^2)^2}{4\gamma c_{\sigma}k_F^3} + \frac
{\gamma c_{\omega}k_F^3}{12\pi^2y^2} + \frac{3}{k_F^3}\int_o^{k_F} dk k^2
\big(\overrightarrow{k}^2 + M^{\star 2}\big)^{1/2}.
\end{equation}

\noindent For pure neutron matter $\gamma=2$ and for nuclear matter
$\gamma=4$. A specification of the coupling constants $c_{\sigma},~
c_{\omega}$ now specifies the equation of state.
\section{Nuclear matter equation of state}
\hspace{0.3in} For nuclear matter, we fix $c_{\sigma}$ and
$c_{\omega}$ by fits to two nuclear matter properties : the
saturation density ($n_s$) and the binding energy
per particle at $\rho_B=n_s$. For these we choose the values 0.153
$fm^{-3}$ and $-16.3$
$MeV$ respectively, as suggested from analysis of experimental data
(M\"oller, Myers, Swiatecki and Treiner \c{59}). This gives

\begin{eqnarray}
c_{\sigma} = 6.2033~fm^2 \nonumber\\
c_{\omega} = 2.9378~fm^2.
\end{eqnarray}

\ni This leads to a value of $0.78M$ for the effective mass of the nucleon in
saturating nuclear matter. The value of $K$ at saturation density that we
get is $\sim$ 700 $MeV$.

\ni Values of the various thermodynamical quantities for nuclear matter
and neutron matter for various densities are presented in Table 3.1a
and Table 3.1b. The quantity $\mu$ appearing
in this table is the chemical potential, given by the relationship :
$\mu =  (P + \varepsilon)/\rho_B$ and $\rho$ is the total mass-energy
density, $\varepsilon/c^2$.
\begin{table}
\cl{\bf TABLE 3.1a}
\vspace{0.2in}
\cl{ EQUATION OF STATE OF DEGENERATE NUCLEAR MATTER}
\cl{ AS GIVEN BY THE PRESENT MODEL}
\vspace{0.2in}
\begin{center}
\begin{tabular}{ccccccc}
\hline
\multicolumn{1}{c}{$k_F$}&
\multicolumn{1}{c}{$\rho_B$}&
\multicolumn{1}{c}{$y$}&
\multicolumn{1}{c}{$\rho$}&
\multicolumn{1}{c}{$P$}&
\multicolumn{1}{c}{$E$}&
\multicolumn{1}{c}{$\mu$}\\
\multicolumn{1}{c}{($fm^{-1}$)}&
\multicolumn{1}{c}{$fm^{-3}$}&
\multicolumn{1}{c}{}&
\multicolumn{1}{c}{$g~cm^{-3}$}&
\multicolumn{1}{c}{$dyn~cm^{-2}$}&
\multicolumn{1}{c}{$MeV$}&
\multicolumn{1}{c}{$MeV$}\\
\hline
1.0 & 0.068 & 0.91 & 1.12 E 14 & $-$1.14 E 33 & 930.98 & 920.42\\
1.1 & 0.090 & 0.87 & 1.49 E 14 & $-$1.81 E 33 & 927.64 & 915.07\\
1.2 & 0.117 & 0.83 & 1.92 E 14 & $-$2.41 E 33 & 924.39 & 912.95\\
1.3 & 0.148 & 0.79 & 2.44 E 14 & $-$4.82 E 33 & 922.52 & 920.49\\
1.4 & 0.185 & 0.75 & 3.05 E 14 &    6.57 E 33 & 924.41 & 946.56\\
1.5 & 0.228 & 0.72 & 3.79 E 14 &    2.23 E 34 & 932.77 & 993.76\\
1.6 & 0.277 & 0.72 & 4.68 E 14 &    4.73 E 34 & 948.91 & 1055.60\\
1.7 & 0.332 & 0.72 & 5.75 E 14 &    8.11 E 34 & 972.47 &1125.00\\
1.8 & 0.394 & 0.74 & 7.04 E 14 &    1.23 E 35 & 1002.30 &1197.90\\
1.9 & 0.463 & 0.76 & 8.57 E 14 &    1.75 E 35 & 1037.30 &1272.40\\
2.0 & 0.540 & 0.78 & 1.04 E 15 &    2.35 E 35 & 1076.30 &1347.60\\
2.1 & 0.625 & 0.80 & 1.25 E 15 &    3.05 E 35 &1118.40 &1423.00\\
2.2 & 0.719 & 0.83 & 1.49 E 15 &    3.86 E 35 &1163.10 &1498.40\\
2.3 & 0.822 & 0.86 & 1.77 E 15 &    4.79 E 35 &1209.70 &1573.80\\
2.4 & 0.934 & 0.88 & 2.09 E 15 &    5.85 E 35 &1257.90 &1649.00\\
2.5 & 1.055 & 0.91 & 2.46 E 15 &    7.05 E 35 &1307.40 &1724.00\\
\hline
\end{tabular}
\end{center}
\end{table}
\vspace {0.2in}
\begin{table}
\cl{\bf TABLE 3.1b}
\vspace{0.2in}
\cl{ EQUATION OF STATE OF DEGENERATE NEUTRON MATTER}
\cl{ AS GIVEN BY THE PRESENT MODEL}
\vspace {0.2in}
\begin{center}
\begin{tabular}{ccccccc}
\hline
\multicolumn{1}{c}{$k_F$} &
\multicolumn{1}{c}{$\rho_B$} &
\multicolumn{1}{c}{$y$} &
\multicolumn{1}{c}{$\rho$} &
\multicolumn{1}{c}{$P$}&
\multicolumn{1}{c}{$E$}&
\multicolumn{1}{c}{$\mu$}\\
\multicolumn{1}{c}{($fm^{-1}$)} &
\multicolumn{1}{c}{$fm^{-3}$} &
\multicolumn{1}{c}{} &
\multicolumn{1}{c}{$g~cm^{-3}$} &
\multicolumn{1}{c}{$dyn~cm^{-2}$}&
\multicolumn{1}{c}{$MeV$}&
\multicolumn{1}{c}{$MeV$}\\
\hline
1.0 & 0.034 & 0.95 & 0.57 E 14 & $-$9.19 E 31 & 941.03 & 939.33\\
1.1 & 0.045 & 0.94 & 0.75 E 14 & $-$2.05 E 32 & 940.39 & 937.53\\
1.2 & 0.058 & 0.92 & 0.98 E 14 & $-$3.76 E 32 & 939.49 & 935.47\\
1.3 & 0.074 & 0.90 & 1.24 E 14 & $-$5.78 E 32 & 938.41 & 933.54\\
1.4 & 0.093 & 0.87 & 1.55 E 14 & $-$6.91 E 32 & 937.32 & 932.67\\
1.5 & 0.110 & 0.84 & 1.90 E 14 & $-$3.56 E 32 & 936.57 & 934.62\\
1.6 & 0.140 & 0.80 & 2.31 E 14 &    1.26 E 33 & 936.82 & 942.49\\
1.7 & 0.170 & 0.77 & 2.78 E 14 &    5.66 E 33 & 939.12 & 960.40\\
1.8 & 0.200 & 0.75 & 3.32 E 14 &    1.46 E 34 & 944.77 & 991.16\\
1.9 & 0.230 & 0.73 & 3.94 E 14 &    2.93 E 34 & 954.84 &1033.80\\
2.0 & 0.270 & 0.73 & 4.67 E 14 &    4.99 E 34 & 969.73 &1084.90\\
2.1 & 0.310 & 0.73 & 5.52 E 14 &    7.60 E 34 & 989.25 &1141.00\\
2.2 & 0.360 & 0.74 & 6.49 E 14 &    1.08 E 35 &1012.90 &1199.90\\
2.3 & 0.410 & 0.75 & 7.62 E 14 &    1.45 E 35 &1040.10 &1260.30\\
2.4 & 0.470 & 0.77 & 8.91 E 14 &    1.88 E 35 &1070.20 &1321.50\\
2.5 & 0.530 & 0.79 & 1.04 E 15 &    2.37 E 35 &1102.70 &1383.20\\
\hline
\end{tabular}
\end{center}
\noindent NOTE.- The respective columns stand for Fermi momentum, nucleon
number density, the nucleon effective mass factor, the total mass-energy,
the pressure, the energy per nucleon and the nucleon chemical potential.
The numbers following the letter $E$ represent powers of ten in all the
tables.
\end{table}
\par
In heavy-ion collision experiments, hot hadronic matter is produced at
temperatures upto 100 $MeV$, which contain upto $25\%$ of their energy in
nuclear resonances and mesonic degrees of freedom. After making allowance
for (model-dependent) corrections for the thermal part of the energy,
first estimates of the energy per nucleon of nuclear matter (for $k_BT = 0$)
have
been made (Stock \c{56} for a discussion). These estimates by Stock
\c{56} are unlikely to be firm, however, for our purpose we have
considered here as guide for comparison with our results. In Fig. 3.1,
we present a
comparison of available such estimates with the prediction of
the equation of state considered by us here. Upto a density of
4$n_s$, there is satisfactory agreement between the two.
\vfill
\begin{figure}
\vskip 7.5in
\noindent {\bf Fig. 3.1}: First generation of estimates, from heavy-ion
collision data, of energy per nucleon, $E$, of nuclear matter plotted
against $n_B=\rho_B$
(in units of $n_s$). The crosses correspond to pion data and the circles
to radial energies (Stock \c{56} for a detailed discussion). The dashed
curve corresponds to the present model (nuclear matter).
\end{figure}
\section{Neutron star matter equation of state}
\hspace{0.3in} At high densities typical of interiors of neutron
stars, the
neutron chemical potentials will exceed the combined masses of proton and
electron. Asymmetric nuclear matter with an admixture of electrons
(rather than pure neutron matter) is, therefore, a more likely composition
of matter in neutron star interiors. The concentrations of protons and
electrons (denoted by $n_p$ and $n_e$ respectively) can be determined using
conditions of beta equilibrium ($n\leftrightarrow p+e+\bar \nu$) and
electrical charge neutrality :

\begin{eqnarray}
\mu_n = \mu_p + \mu_e  \nonumber\\
n_p = n_e
\end{eqnarray}
($\mu_i$ = chemical potential of particle species $i$).
\par
Since nuclear force is known to favour isospin symmetry, and since the
symmetry energy arising solely from the Fermi energy is known to be
inadequate to account for the empirical value of the symmetry energy
($\simeq$ 32 $MeV$), we include the interaction due to isospin triplet $\rho$-
meson in Eq. (3.1) for purpose of describing neutron-rich matter. That is,
we add the following terms :

\begin{equation}
-\frac {1}{4}G_{\mu\nu}G^{\mu\nu}
+\frac{1}{2}m^2_{\rho}\overrightarrow{\rho_{\mu}}.
\overrightarrow{\rho}^{\mu}
-\frac{1}{2}g_{\rho}\bar\psi(\overrightarrow{\rho}_{\mu}.
\overrightarrow{\tau}\gamma^{\mu}) \psi
\end{equation}

\noindent to the right hand side of Eq. (3.1) in order to describe the
asymmetric matter. Here, $\overrightarrow{\rho}_{\mu}$ stands for the
$\rho$- meson field with mass $m_{\rho}$, $g_{\rho}$ is the coupling
strength and

\begin{equation}
G_{\mu\nu} \equiv \partial_{\mu}\overrightarrow{\rho}_{\nu}-\partial_{\nu}
\overrightarrow{\rho}_{\mu}.
\end{equation}

\ni Strictly speaking, the $\rho$- meson should couple to the total conserved
current (Glendenning, Banerjee and Gyulassy \c{60}). In the above, we
have coupled the $\rho$- meson to the baryons, which are not the only
possible source of isospin current. However, for the ground state
equation of state, in the mean-field approximation, only the baryon part of the
isospin current will survive (Glendenning \c{40}).

\noindent The equation of motion for $\overrightarrow{\rho}_{\mu}$, in the mean
field approximation where $\overrightarrow{\rho}_{\mu}$ is replaced by its
uniform value $\rho_o^3$ (here superscript 3 stands for the third
component in isospin space), gives

\begin{equation}
m^2_{\rho} \rho^3_o = \frac{1}{2}g_\rho\sum_{B=n,p}<\bar{\psi}\gamma_o
\tau^3\psi >_B,
\end{equation}

\noindent where the sum is over neutrons and protons. This gives the following
density dependence for the field variable $\rho^3_o$ :

\begin{equation}
\rho^3_o = \frac{g_{\rho}}{2m_\rho^2} (n_p-n_n).
\end{equation}

\ni The symmetric energy coefficient that follows from the semi-empirical
nuclear mass formula (that is, the coefficient of the term $(n_p
- n_n)^2/(n_p + n_n)^2$ in the mass formula), is:

\begin{equation}
a_{sym} = \frac{c_{\rho} k_F^3}{12\pi^2} + \frac{k_F^2}{6(k_F^2
+M^{\star 2})^{1/2}},
\end{equation}

\ni where $c_{\rho} \equiv g^2_\rho/m^2_{\rho}$ and $k_F=(6\pi^2
\rho_B/\gamma)^{1/3}$ ($\rho_B~=~n_p+n_n$) . We fix the coupling constant
$c_{\rho}$ by requiring that $a_{sym}$  correspond to the empirical value
32 $MeV$. This gives

\begin{equation}
c_{\rho} = 4.6617~fm^2.
\end{equation}

\ni It is noted that the $\rho$- meson will contribute a term =
$m^2_{\rho}(\rho_o^3)^2/2$ to the energy density and pressure.
Table 3.2 lists the pressure versus the total mass-energy density for the
neutron-rich matter in beta equilibrium.
\begin{table}
\cl{\bf TABLE 3.2}
\vspace {0.2in}
\cl{PRESSURE VS DENSITY FOR NEUTRON-RICH MATTER}
\vspace {0.2in}
\begin{center}
\begin{tabular}{cc}
\hline
\multicolumn{1}{c}{$\rho$} &
\multicolumn{1}{c}{$P$} \\
\multicolumn{1}{c}{($g~cm^{-3}$)} &
\multicolumn{1}{c}{($dyn~cm^{-2}$)}\\
\hline
4.531 E 15 & 1.619 E 36 \\
3.686 E 15 & 1.285 E 36 \\
2.695 E 15 & 8.978 E 35 \\
2.109 E 15 & 6.719 E 35 \\
1.784 E 15 & 5.472 E 35 \\
1.504 E 15 & 4.406 E 35 \\
1.211 E 15 & 3.296 E 35 \\
1.017 E 15 & 2.565 E 35 \\
9.116 E 14 & 2.171 E 35 \\
8.175 E 14 & 1.822 E 35 \\
7.028 E 14 & 1.400 E 35 \\
6.059 E 14 & 1.051 E 35 \\
5.576 E 14 & 8.813 E 34 \\
5.037 E 14 & 6.967 E 34 \\
4.561 E 14 & 5.410 E 34 \\
4.063 E 14 & 3.887 E 34 \\
3.564 E 14 & 2.534 E 34 \\
3.018 E 14 & 1.352 E 34 \\
\hline
\end{tabular}
\end{center}
\end{table}
\section{Results}
\hspace{0.3in} Using the chiral sigma model and adopting the
approach that the isoscalar
vector field, needed to provide saturating binding energy of degenerate
nuclear matter be generated dynamically, we have obtained an
equation of state of
degenerate nuclear and neutron-rich matter at high densities.
\par
The maximum mass of the neutron stars is calculated by integrating
the structure equations, which will be discussed in the chapter 7 in
detail. A comparison with the maximum mass of (nonrotating)
neutron stars predicted
by others, using models based on recent field theoretical equations of
state is
given in Table 3.3. The choice of the equation-of-state models
in Table 3.3 is representative, but by no means exhaustive.
Included in this comparison are equation of state models due to Alonso and
Cabanell \c{36} and Prakash and Ainsworth \c{41} which are also based
on the sigma
model, but differ from our model in the details. The equation of
state (II) of Alonso and
Cabanell \c{36}, comes from a determination of the free
parameters
of the linear sigma model with an explicit symmetry-breaking term and is
coupled to $\omega$- and $\rho$- mesons, in a renormalizable way.
Prakash and Ainsworth \c{41} included the sigma meson
one-loop contributions, but the isoscalar vector field was not generated
dynamically, so that its role is reduced to an empirical one, allowing for
arbitrary variations in its coupling constant (thereby making it possible
to obtain any desired value of the nuclear matter compression
modulus). The vector field plays no role in determining the value of the
effective mass of the nucleon in such an approach. A comparison
of this equation of state
with our model for neutron-rich matter and pure neutron matter is shown in
Fig. 3.2. The equation of state of Serot \c{61} is calculated in the mean-field
approximation in the Walecka model including $\sigma-,~ \omega$- and $\rho$-
mesons and that of Chin \c{62} with one-loop corrections in the
$\sigma-\omega$  model. The equation of state
due to Glendenning \c{40} is along similar lines, and includes one-loop
corrections and also scalar self-interactions (upto quartic order), whose
magnitudes are adjusted to reproduce empirical saturation properties.
This work takes into account the effect of hyperons in beta equilibrium in
addition to electrons and muons.
\begin{table}
\cl{\bf TABLE 3.3}
\vspace{0.2in}
\cl {MAXIMUM MASS NEUTRON STARS FOR FIELD THEORETICAL}
\cl{ EQUATION OF STATE MODELS }
\vspace {0.2in}
\begin{center}
\begin{tabular}{ccccc}
\hline
\multicolumn{1}{c}{$Equation~ of~ state~ Reference$} &
\multicolumn{1}{c}{$n_s$}&
\multicolumn{1}{c}{$K$}&
\multicolumn{1}{c}{$M_{max}/M_{\odot}$}&
\multicolumn{1}{c}{$R$}\\
\multicolumn{1}{c}{} &
\multicolumn{1}{c}{$fm^{-3}$} &
\multicolumn{1}{c}{$MeV$} &
\multicolumn{1}{c}{}&
\multicolumn{1}{c}{$km$}\\
\hline
Serot \c{61} & 0.193 & 540 & 2.54 & 12.28 \\
Chin \c{62}  & 0.193 & 471 & 2.10 & 10.57 \\
Alonso-Cabanell \c{36} equation of state II & 0.172 &225 & 1.94 & 10.90\\
Prakash-Ainsworth \c{41} (g$^2_{\omega}$ = 16.27) & 0.160 & 225 &
1.83 & 9.89 \\
Glendenning \c{34} & 0.153 & 300 & 1.79 & 11.18 \\
Present model \c{54a} & 0.153 & 700 & 2.59 & 14.03 \\
\hline
\end{tabular}
\end{center}
\noindent NOTE.- n$_s$ and $K$ are nuclear matter saturation density and
compression modulus.
\end{table}
\vspace {0.2in}
\begin{figure}
\vskip 7.5in
\noindent{\bf Fig. 3.2}: Pressure versus total mass-energy density. Curve 1
corresponds to neutron-rich matter in beta equilibrium (including the
contribution from $\rho-$ meson exchange) : present model. Curve 2
corresponds to pure neutron matter : present model. Curve 3 (dashed curve)
is due to Prakash and Ainsworth \c{41}.
\end{figure}
\par
Table 3.3 implies that the present
neutron-rich matter equation of state is comparatively `stiff'
as far as neutron
stars are concerned. This is reflected in the value
of the maximum mass of neutron stars, which is the largest for the present
model. It may be mentioned here that observational evidence in favour of
a stiff equation of state comes from the identification by
Tr\"umper $et~al.$ \c{64} of the
35 day cycle of the pulsating X-ray source Her X-1 as originating in free
precession of the rotating neutron star (Pines \c{65}).
\par
In constructing the neutron star matter equation of state, we
have restricted ourselves
to (n, p, e$^{-}$) matter in beta equilibrium. At high densities,
hyperons can also appear
(Glendenning \c{34}; Kapusta and Olive \c{35}); this is expected
to reduce the
`stiffness' in the equation of state, and a consequent reduction
in the maximum mass of
neutron stars.
Another point that we have not investigated here is the possible
role of the $\rho-$ meson tensor interaction as far as the symmetry energy
is concerned.
\vfill
\newpage
\setcounter{equation}{0}
\chapter{Astrophysical application ~ I : Structure and radial oscillation
of nonrotating neutron stars}
\section{ The Neutron Star Structure}
\hspace {0.3in} The structure of a neutron star is characterized
by its mass and radius.
Additional parameters of interest are the moment of inertia and the crust
thickness. These are important for the dynamics and transport properties
of pulsars.
\par
The space-time for a spherically symmetric gravitating system is
described by the Schwarzschild metric

\begin{equation}
ds^2 = e^{\nu} c^2 dt^2 - r^2 (d\theta^2 + sin^2\theta d\phi^2) -
e^{\lambda} dr^2,
\end{equation}

\noindent where $\nu,~\lambda$ are functions of $r$ only \c{114}. Corresponding
to this space-time metric, the equations that describe the hydrostatic
equilibrium of degenerate stars without rotation in general
relativity ($i.e.$, where
temperature and convection are not important) can be written as \c{114} :

\begin{equation}
\frac {dp}{dr} = - \frac{G (\rho + p/c^2) (m + 4\pi r^3 p/c^2)} {r^2
(1-2 Gm/rc^2)}
\end{equation}

\begin{equation}
\frac {dm}{dr} = 4\pi r^2\rho ,
\end{equation}

\noindent where $p$ and $\rho$ are the pressure and total mass energy
density and $m(r)$ is the mass contained in a volume of radius $r$. Given
an equation of state p($\rho$), Eqs. (7.2) and (7.3), called
Tolman-Oppenheimer-Volkoff (TOV) equation, can be numerically integrated, for a
given central density, to obtain the radius $R$ and gravitational mass $M_G =
m(R)$ of the star.
\par
The moment of inertia $I$ of the rotating neutron star, rotating
with angular velocity $\Omega$ as seen by a distant observer is
given by \c{115}

\begin{eqnarray}
I = \frac{1}{\Omega} \frac{c^2 R^4}{6G} \Big(\frac
{d\bar{\omega}}{dr}\Big)_{r=R},
\end{eqnarray}

\noindent where $\bar{\omega}(r)$, the angular velocity of the star fluid
relative to the local inertial frame, is given by

\begin{equation}
\frac{d}{dr} \Big(r^4j\frac{d\bar{\omega}}{dr}\Big) +
4r^3\bar{\omega}\frac{dj}{dr} = 0,
\end{equation}

\noindent where

\begin{equation}
j(r) = e^{-\nu/2} \Big(1-2GM_G/rc^2\Big)^{1/2}
\end{equation}

\noindent and satisfies the boundary conditions

\begin{equation}
\Big(\frac{d\bar{\omega}}{dr}\Big)_{r=R} = 0;\; \bar{\omega} (\infty) =
\Omega.
\end{equation}
\noindent Here the potential  function $\nu(r)$, relating the element of
proper time to the element of time at $r=\infty$ is given by

\begin{eqnarray}
\frac{d\nu}{dr} = \frac{2G}{r^2c^2} \frac{(m + 4\pi r^3
P/c^2)}{(1-2Gm/rc^2)}.
\end{eqnarray}

\ni The definition (7.4) for $I$ includes the relativistic Lens-Thirring
effect to order $\Omega^{2}$. For neutron star, the relativistic effect
are important, and so one gets $I$ as defined by Eq. (7.4) to be in excess
of classical definition $(2/5)MR^2$.
\par
For the numerical integration to obtain the structure parameters, it
is sufficient to start with an arbitrary value of $\nu(0)$, which is
then rescaled to satisfy the surface condition

\begin{equation}
\nu(R) = ln \Big(1 - \frac{2GM_G}{Rc^2}\Big),
\end{equation}

\noindent so that $\nu(\infty) = 0$. Likewise, $\bar{\omega}(0)$ is
initially chosen to be an arbitrary constant, and a value of $\Omega$
given by

\begin{equation}
\Omega = \bar{\omega}(R) + \frac{1}{3} R
\Big(\frac{d\bar{\omega}}{dr}\Big)_{r=R}
\end{equation}

\noindent obtained. A new starting value $\bar{\omega}_{new}(0)$
corresponding to any specified $\Omega_{new}$ is given by

\begin{equation}
\bar{\omega}_{new}(0) = \Omega_{new} \bar{\omega}(0)/\Omega.
\end{equation}
\par
To integrate the TOV equations, one needs to know the equation of state
$p(\rho)$, for the entire expected density range of neutron star,
starting from the
high density at the center to the surface densities. The composite equation
of state for the entire neutron star density span, was constructed by
joining the equation of state of high density neutron-rich matter
(that we have discussed in chapter 3) to that given by (a) Negele and
Vautherin \c{116} for
the density region ($10^{14} - 5 \times 10^{10}$) $g~ cm^{-3}$, (b) Baym,
Pethick and Sutherland \c{117} for the region ($5 \times 10^{10} - 10^{3}$) $g~
cm^{-3}$ and (c) Feynman, Metropolis and Teller \c{118} for $\rho < 10^3$ $g~
cm^{-3}$. These densities regions are tabulated in Table 7.1--7.3.
\begin{table}
\cl{\bf TABLE 7.1}
\vspace {0.2in}
\cl {PRESSURE VS DENSITY FOR NEGELE AND VAUTHERIN}
\cl{EQUATION OF STATE}
\vspace {0.2in}
\begin{center}
\begin{tabular}{cc}
\hline
\multicolumn{1}{c}{$\rho$} &
\multicolumn{1}{c}{$P$} \\
\multicolumn{1}{c}{($g~cm^{-3}$)} &
\multicolumn{1}{c}{($dyn~cm^{-2}$)}\\
\hline
1.586 E 14 & 8.617 E 32 \\
9.826 E 13 & 3.807 E 32 \\
6.193 E 13 & 1.835 E 32 \\
3.767 E 13 & 8.564 E 31 \\
2.210 E 13 & 3.789 E 31 \\
1.496 E 13 & 2.095 E 31 \\
9.611 E 12 & 1.095 E 31 \\
6.248 E 12 & 6.184 E 30 \\
3.833 E 12 & 3.621 E 30 \\
2.202 E 12 & 2.276 E 30 \\
1.471 E 12 & 1.694 E 30 \\
9.728 E 11 & 1.228 E 30 \\
6.610 E 11 & 8.633 E 29 \\
5.228 E 11 & 6.741 E 29 \\
\hline
\end{tabular}
\end{center}
\end{table}
\begin{table}
\cl{\bf TABLE 7.2}
\vspace {0.2in}
\cl {PRESSURE VS DENSITY FOR BAYM, PETHICK AND SUTHERLAND}
\cl{EQUATION OF STATE}
\vspace {0.2in}
\begin{center}
\begin{tabular}{cc}
\hline
\multicolumn{1}{c}{$\rho$} &
\multicolumn{1}{c}{$P$} \\
\multicolumn{1}{c}{($g~cm^{-3}$)} &
\multicolumn{1}{c}{($dyn~cm^{-2}$)}\\
\hline
5.254 E 10 & 5.949 E 28 \\
3.313 E 10 & 3.404 E 28 \\
2.090 E 10 & 1.938 E 28 \\
1.318 E 10 & 1.048 E 28 \\
8.312 E 09 & 5.662 E 27 \\
4.164 E 09 & 2.356 E 27 \\
1.657 E 09 & 6.860 E 26 \\
1.045 E 09 & 4.129 E 26 \\
5.237 E 08 & 1.629 E 26 \\
2.624 E 08 & 6.676 E 25 \\
6.589 E 07 & 1.006 E 25 \\
1.655 E 07 & 1.435 E 24 \\
6.588 E 06 & 3.911 E 23 \\
1.044 E 06 & 2.318 E 22 \\
1.654 E 05 & 1.151 E 21 \\
2.622 E 04 & 4.968 E 19 \\
1.044 E 04 & 9.744 E 18 \\
\hline
\end{tabular}
\end{center}
\end{table}
\begin{table}
\cl{\bf TABLE 7.3}
\vspace {0.2in}
\cl {PRESSURE VS DENSITY FOR FEYNMAN, METROPOLIS AND TELLER }
\cl{EQUATION OF STATE}
\vspace {0.2in}
\begin{center}
\begin{tabular}{cc}
\hline
\multicolumn{1}{c}{$\rho$} &
\multicolumn{1}{c}{$P$} \\
\multicolumn{1}{c}{($g~cm^{-3}$)} &
\multicolumn{1}{c}{($dyn~cm^{-2}$)}\\
\hline
2.120 E 03 & 5.820 E 15 \\
4.500 E 01 & 1.700 E 14 \\
1.640 E 01 & 1.400 E 13 \\
1.160 E 01 & 1.210 E 12 \\
8.150 E 00 & 1.010 E 11 \\
7.900 E 00 & 1.010 E 10 \\
\hline
\end{tabular}
\end{center}
\end{table}
\par
For a given equation of state $P(\rho)$, and a given central density
$\rho(r=0) = \rho_c$, the Eqs. (7.2-7.3) are integrated numerically with
the boundary condition :

\begin{eqnarray}
m(r = 0) = 0
\end{eqnarray}

\noindent to give $R$ and $M_G$. The radius $R$ is defined by the point
where $P$ $\simeq$ 0, or, equivalently, $\rho = \rho_s$, where $\rho_s$
is the density expected at the neutron star surface (about 7.8
$g~cm^{-3}$). The total gravitational mass is then given by :
$M_G = m(R)$.
\par
We solved \c{54a} the TOV equation numerically using
predictor-corrector method, which gives better accuracy compared
to existing results of the gravitational mass and the radius.
In this method, we use logarithm of pressure and energy density
$e.g.$, equation of state to take care of lower energy density and
pressure, because at low densities the energy density falls very rapidly
at the surface of the star.
\par
Table 7.4 lists neutron star structure parameters as predicted by our
equation of state (chiral sigma model) for
neutron-rich matter, which has already been discussed in chapter 3.
The maximum gravitational mass for stable
non-rotating neutron star predicted by our model is 2.59 $M_{\odot}$
($M_{\odot}$ = solar mass). This occurs for a central density of
$1.4 \times 10^{15}$ $g~cm^{-3}$.
The corresponding radius and entire crustal length $\Delta$
of the star are 14.03 $km$ and 1 $km$ respectively. The crust length
$\Delta$ is defined as the distance overwhich the density falls from
$\rho = 2.4\times 10^{14}$ to 7.8 $g~~cm^{-3}$ $i.e.,$ the surface.
Fig. 7.1 and Fig. 7.2 show the plots of
mass vs central density and moment of inertia vs mass.
The maximum moment of inertia is $4.79 \times 10^{45}$ $g~cm^2$.

\begin{table}
\cl{\bf TABLE 7.4}
\vspace{0.2in}
\cl{NEUTRON STAR STRUCTURE PARAMETERS FOR THE PRESENT MODEL }
\cl{NEUTRON STAR EQUATION OF STATE }
\vspace {0.2in}
\begin{center}
\begin{tabular}{cccccc}
\hline
\multicolumn{1}{c}{$\rho_c$} &
\multicolumn{1}{c}{$R$}&
\multicolumn{1}{c}{$M/M_{\odot}$}&
\multicolumn{1}{c}{$I$}&
\multicolumn{1}{c}{$\alpha$}&
\multicolumn{1}{c}{$\Delta$}\\
\multicolumn{1}{c}{($g~cm^{-3}$)} &
\multicolumn{1}{c}{($km$)} &
\multicolumn{1}{c}{} &
\multicolumn{1}{c}{($g~cm^{2}$)}&
\multicolumn{1}{c}{}&
\multicolumn{1}{c}{($km$)}\\
\hline
3.5 E 14 & 14.30 & 1.02 & 1.31 E 45 & 0.889&4.06\\
4.0 E 14 & 14.73 & 1.36 & 2.05 E 45 & 0.853&3.11\\
4.5 E 14 & 14.99 & 1.64 & 2.73 E 45 & 0.822&2.55\\
5.0 E 14 & 15.13 & 1.86 & 3.30 E 45 & 0.798&2.18\\
6.0 E 14 & 15.18 & 2.16 & 4.07 E 45 & 0.761&1.76\\
8.0 E 14 & 14.94 & 2.45 & 4.70 E 45 & 0.719&1.36\\
1.0 E 15 & 14.62 & 2.55 & 4.79 E 45 & 0.696&1.18\\
1.2 E 15 & 14.30 & 2.59 & 4.70 E 45 & 0.682&1.07\\
1.4 E 15 & 14.03 & 2.59 & 4.54 E 45 & 0.674&1.00\\
\hline
\end{tabular}
\end{center}
\noindent NOTE.- The respective columns stand for central density, radius,
gravitational mass, moment of inertia calculated for angular velocity =
($GM_G/R^3)^{1/2}$, and the surface  red shift ratio $\alpha$.
\end{table}

\begin{figure}
\vskip 7.5in
\noindent{\bf Fig. 7.1}: Gravitational mass ($M_G$) of non-rotating neutron
stars versus central density ($\rho_c$) as predicted by the present model
equation of state (neutron-rich matter). The maximum stable mass is 2.59
$M_{\odot}$.
The corresponding central density, $\rho_c$ = 1.5 $\times$ 10$^{15}$ $g~
cm^{-3}$ and the radius is 14.0 $km$.
\end{figure}

\begin{figure}
\vskip 7.5in
\noindent{\bf Fig. 7.2}: Neutron star moment of inertia ($I$) versus
gravitational mass ($M_G$), as predicted by the present model.
\end{figure}
\par
Observationally, masses of neutron stars are estimated from compact binary
systems, one member of which is a pulsar. The most precise estimate comes
from the pulsar PSR 1913+16, which gives (1.442 $\pm$0.003) $M_{\odot}$.
A recent compilation of the estimated masses by X-ray pulsars (Nagase
\c{119}) gives the maximum mass (corresponding to Vela X-1 pulsar) to be
(1.77 $\pm$0.21) $M_{\odot}$. Stable neutron star masses predicted by the
present equation of state discussed in chapter 3 are thus compatible with the
observational estimates. The
surface  red shift factor provides a probe for neutron star structure, if
one presumes that observed $\gamma-$ray bursts are gravitationally
 red shifted $e^+e^-$ annihilation lines, produced near their surface. The
surface  red shift ratio ($\alpha$) defined as

\begin{equation}
\alpha = (1 - 2GM_G/Rc^2)^{1/2}
\end{equation}

\ni is expected to be (0.78 $\pm$0.02) on the basis of observed data
(Friedman and Pandharipande \c{27} for a discussion). The present
neutron-rich matter equation of state, which has already been discussed,
gives for a 1.4 $M_{\odot}$ neutron star :  $R$ = 14.77 $km$,
$I$ = 2.15 $\times$
10$^{45}$ $g~cm^2$ $\Delta$=3.0 $km$ and the  red shift ratio (at
the surface) = 0.85. The corresponding central density is 4.06
$\times$ 10$^{14}$ $g~cm^{-3}$.

\section{Neutron stars radial oscillations}
\hspace{0.3in} Since the original suggestion by Cameron \c{120}
that vibration of neutron
star could excite motions, which might have interesting astrophysical
applications, there has been a series of investigations into the
vibrational properties of neutron stars. The earliest detailed
calculations, carried out by Meltzer and Thorne \c{121} and Thorne \c{122}
examined the radial as well as nonradial oscillations of neutron stars,
using then available equation of state, such as the
Harrison-Wakono-Wheeler \c{122a}  equation of state. These and other early
studies,
$e.g.$, Wheeler \c{123}; Chau \c{124} and Occhionero \c{125},
indicated that the
fundamental mode radial oscillation periods for the neutron stars would
typically lie in the vicinity of about 0.4 $ms$, and that the first few
quadrupole oscillations would have periods that are also fractions of a
millisecond. Furthermore, these oscillation periods were estimated to be
damped by gravitational radiation with damping time scales of the order of
one second. Also, there are recent papers on nonrotating, rapidly rotating
neutron stars and slowly rotating stars. For example, in 1990,
Cutler, Lindblom  and Splinter \c{126} computed the frequencies and
damping times due to viscosity and gravitational radiation for
the lowest frequency modes of a wide range of nonrotating fully
relativistic neutron star models. In a subsequent paper,
Cutler and Lindblom \c{127} reviewed and extended the formalism for
computing the oscillation frequency of rapidly rotating neutron
stars in the post-Newtonian approximation. Recently, Kojima
\c{128} calculated the rotational shift of normal frequencies in
polytropic stellar models in the framework of general
relativity. The stellar rotation is assumed to be slow and
first-order rotational effects are included to the
eigenfrequencies of the nonrotating stars. In this section, we
present the calculation of
radial oscillation  periods of neutron stars, using our equation
of state based on the chiral sigma model.
\par
The equation governing infinitesimal radial pulsations of a
nonrotating star in general relativity was given by Chandrasekhar
\c{129}, and it has the following form :

\begin{equation}
F \frac {d^2\xi}{dr^2} + G \frac{d\xi}{dr} + H\xi = \sigma^2\xi ,
\end{equation}

\noindent where $\xi$(r) is the Lagrangian fluid displacement and
$c\sigma$ is the characteristic eigenfrequency ($c$ is the velocity
of light). The quantities $F$, $G$,
$H$ depend on the equilibrium profiles of the pressure and density of
the star, and are given by

\begin{equation}
F = - e^{-\lambda} e^{\nu} \Gamma p/(p+\rho c^2)
\end{equation}

\begin{eqnarray}
G = - e^{-\lambda} e^{\nu} \Bigl\{\Gamma p \Big(\frac {1}{2}
\frac{d\nu}{dr} + \frac{1}{2} \frac{d\lambda}{dr} +
\frac{2}{r}\Big) + \nonumber \\
p \frac{d\Gamma}{dr} + \Gamma \frac{dp}{dr}\Bigr\}/(p+\rho c^2)
\end{eqnarray}

\begin{equation}
H = \frac {e^{-\lambda}e^{\nu}}{p + \rho c^2} \Bigl\{ \frac{4}{r}
\frac{dp}{dr} - \frac{(dp/dr)^2}{p + \rho c^2} - A \Bigr\} + \frac
{8\pi G}{c^4} e^{\nu} p ,
\end{equation}

\noindent where $\Gamma$ is the adiabatic index, defined in the
general relativistic case as

\begin{equation}
\Gamma = (1 + \rho c^2/p) \frac{dp}{d(\rho c^2)}
\end{equation}

\noindent and

\begin{eqnarray}
A = \frac {d\lambda}{dr} \frac{\Gamma p}{r} + \frac
{2p}{r} \frac{d\Gamma}{dr} + \frac {2\Gamma}{r} \frac{dp}{dr} -
\frac{2\Gamma p}{r^2}\nonumber\\
- \frac{1}{4} \frac{d\nu}{dr} \Big( \frac{d\lambda}{dr} \Gamma p +
2p \frac{d\Gamma}{dr} + 2\Gamma \frac{dp}{dr} - \frac{8\Gamma
p}{r}\Big)\nonumber\\ - \frac{1}{2} \Gamma p
\Big(\frac{d\nu}{dr}\Big)^2 - \frac{1}{2}
\Gamma p \frac{d^2\nu}{dr^2}.
\end{eqnarray}

\ni The boundary conditions to solve the pulsation equation (7.14) are

\begin{equation}
\xi (r = 0) = 0
\end{equation}

\begin{equation}
\delta p (r = R) = -\xi \frac{dp}{dr} - \Gamma p \frac
{e^{\nu/2}}{r^2} \frac{\partial}{\partial r} (r^2 e ^{-\nu/2}
\xi)\vert_{r=R} = 0
\end{equation}

\ni Here $\Gamma$ is the adiabatic index.
Since $p$ vanishes at $r=R$, it is generally sufficient to demand

\begin{equation}
\xi~~ finite~~ at~~ r=R.
\end{equation}

\noindent Equation (7.14), subject to the boundary conditions Eq.
(7.20) and (7.21), is a Sturm - Liouville eigen value equation
for $\sigma^2$.

\ni The following results follow from the theory of such equations :

\begin{enumerate}

\item The eigen values $\sigma^{2}$ are all real.

\item The eigen values from an infinite discrete sequence

\begin{eqnarray}
\sigma^2_o < \sigma^2_1 <\dots < \sigma^2_n <\dots\dots,
\end{eqnarray}

\item The eigen function $\xi_o$ corresponding to $\sigma_o^{2}$ has no
nodes (fundamental mode) in the interval $0<r<R$; more generally,
$\xi_n$ has nodes in this interval.

\item  The $\xi_n$ are orthogonal with weight function $\omega r^2$ :
$$\int_{0}^{R} \xi_n \xi_m \omega r^2 dr =0,~ m \neq n.$$

\item  The $\xi_n$ form a complete set for the expansion of any function
satisfying the boundary condition Eqs. (7.20) and (7.22).
\end{enumerate}

\ni An important consequence of item (2) is the following :

\ni If the fundamental radial mode of a star is stable ($\sigma_o^2 > 0$), then
all the radial modes are stable. Conversely, if the star is radially
unstable, the fastest growing instability will be via the fundamental
mode ($\sigma_o^{2}$ more negative than all other $\sigma_n^{2}$).
\par
We solved \c{129a} Eq. (7.14) for the eigenvalue $\sigma$ by writing the
differential equation as a set of difference equations. The
equations were cast in tridiagonal form and the eigenvalues were found by
using the EISPACK routine. This routine finds the eigenvalues of a
symmetric tridiagonal matrix by the implicit QL method.
\begin{figure}
\vskip 7.5in
\noindent{\bf Fig. 7.3}: Periods of radial pulsations as functions of the
gravitational mass for our equation of state (chiral sigma
model). The labels 1, 2, 3, 4, 5 correspond respectively to the
fundamental and the first four harmonics.
\end{figure}
\par
Results for the oscillations of neutron star corresponding to
chiral sigma equation of state are illustrated in Fig. 7.3.
The plot in Fig. 7.3 are for the oscillation time period (=
2$\pi/c\sigma$) versus the gravitational mass $M_G$. The
fundamental mode and the first four harmonics are considered.
The period is an increasing function of $M_G$,
the rate of increase being progressively less for higher oscillation
modes. The fundamental mode  oscillation periods for neutron star are
found to have the following range of values : (0.35 - 0.50) $milliseconds$.
For higher modes, the periods are $\leq$ 0.2 $milliseconds$. We
shall compare our results with those for a different choice of
the equation of state, namely, that given by Wiringa $et~al.$ \c{10}.
We also perform a similar calculation for strange quark stars,
using a realistic equations of state.

\section{Radial oscillations of quark stars}
\hspace{0.3in} There are strong reason for
believing that the hadrons are composed of quarks, and the idea of
quark stars has already existed for about twenty years \c{130,131}.
Calculations of the possible phase transition from baryon matter to
quark matter in models of cold, compact stars have been performed by
several groups (\c{93,132,94,134}), but the results are not
conclusive concerning the existence of quark matter inside neutron
stars. In 1984, it was suggested that strange matter $i.e.$, quark
matter with strangeness per baryon of order unity, may be the true ground
state \c{97}. The properties of strange matter at zero pressure
were subsequently examined, and it was found that strange matter can
indeed be stable for a wide range of parameters in the strong
interaction calculations \c{95}. Details of the extension to finite
pressure and the so-called strange stars are given in \c{96,135}. The
problem of the existence of strange stars is, however, still
unresolved (\c{98}).
\par
An important question is how can one possibly distinguish
between quark stars and neutron stars. It has been suggested to
use measurements of the surface gravitational
red shift $z$ (Schwarzschild), since different equation of state
gives different results for $z(M_G)$, $M$ being the star mass
\c{136,137}. The region of allowed high density equation of state
may be narrowed further by the observations of pulsar periods
\c{138,139}. Given a sub-millisecond pulsar, we may argue that the
ability of such a fast rotating star to avoid rotational
break-up induces severe restrictions and a conventional neutron
star will not be able to resist the large centrifugal forces.
The problem of rapid rotation of compact stars receives much
attention \c{140,141,142,143}, and although no sub-millisecond pulsar is
seen among the about 500 pulsars observed so far, further
observations may well reveal such an object. Also, among the
criteria suggested for  distinguishing quark star from neutron
star are the neutrino cooling rate \c{144,145,146}, transport properties such
as bulk viscosity \c{147,148}, and sub millisecond period rotation
rates \c{141}.
\par
These so-called strange stars have rather different
mass-radius relationship \c{135} than neutron stars, but for stars of
mass = 1.4 $M$$_{\odot}$, the structure parameters of quark stars
are very similar to those of neutron stars. Since pulsars are
believed to be (rotating) neutron stars, and since available
binary pulsar data suggest their masses to be close to 1.4
$M$$_{\odot}$, it has been conjectured \c{97} that at least some
pulsars could be quark stars.
\par
Recently, Haensel $et~al.$ \c{149} have emphasized that
pulsation properties of a neutron star can yield information about
the interior composition, namely, whether the interior has undergone a
phase transition to quarks. The main idea is to know the damping
times, which will be modified if there is a quark matter core.
In their study, Haensel $et~al.$ \c{149} used polytropic model for the
equation of state  for nuclear matter as well as quark matter,
and the Newtonian pulsation equation to calculate the
eigenfrequencies. The strange quark mass and the quark interactions
are important for the structure of quark stars \c{97}. This
suggests that the equation of state of strange quark matter will have a role
to play in determining the pulsation features of quark stars.
Clearly, for a more exact understanding of the vibrational
properties of quark stars, use of realistic equation of state for quark
matter, and the general relativistic pulsation equation, are desirable.
Cutler $et~al.$ \c{126} have calculated the frequencies and damping times
of radial pulsations of some quark star configurations, using the
general relativistic pulsation equation, but for quark matter,
they adopted the MIT bag model in its simplest form, namely,
non-interacting and massless quarks. The purpose of this work \c{129a}
is to calculate the range of eigenfrequencies of radial
pulsations of stable quark stars (using the general relativistic
pulsation equation) and to investigate the sensitivity of the
eigenfrequencies on the equation of state.
\par
The equation of state used by us incorporates short-range
quark-gluon interactions perturbatively to second order in the
coupling constant $\alpha_c$. The long-range interactions are
taken into account phenomenologically by the bag pressure term $B$.
We incorporate the density dependence of $\alpha_c$ by solving the
Gell-Mann-Low equation for the screened charge. The parameters
involved are the strange quark mass $m_s$, $B$ and, the renormalization
point $\mu_o$, obtained by demanding that the bulk strange matter be
stable at zero temperature and pressure, with energy per baryon less
than the lowest energy per baryon found in nuclear matter.
For completeness, we also do the calculations for the MIT bag model.
\par
At high baryonic densities, bulk strange matter is in an overall
colour singlet state, and can be treated as a relativistic Fermi gas
interacting perturbatively. The quark confinement property is being
simulated by the phenomenological bag model constant $B$. Chemical
equilibrium between the three quark flavours and electrical charge
neutrality allow us to calculate the equation of state from the thermodynamic
potential of the system as a function of the quark masses, the bag
pressure term $B$ and the renormalization point $\mu_o$. To second order
in $\alpha_c$, and assuming $u$ and $d$ quarks to be massless, the
thermodynamic potential is given by \c{150} :

\begin{equation}
\Omega = \Omega_u + \Omega_d + \Omega_s + \Omega_{int.} +
\Omega_e,
\end{equation}

\noindent where $\Omega_i$ ($i = u,~ d,~ s,~ e$) represents the
contributions of
$u,~ d,~ s$ quarks and electrons  and
$\Omega_{int}$ is the contribution due to interference between $u$ and $d$
quarks and is of order $\alpha^2_c$ :
Expressions for $\Omega_{i}$ and $\Omega_{int}$ are already given in
chapter 5.

\ni The total energy density and the external pressure of the system are
given by

\begin{equation}
\epsilon  = \Omega + B + \sum_i \mu_i n_i
\end{equation}
\ni and
\begin{equation}
p  = - \Omega - B,
\end{equation}

\noindent where $n_i$ is the number density of the $i$-th particle
species. For specific choices of the parameters of the theory
(namely, $m_s$, $B$ and $\mu_o$), the equation of state is now obtained by
calculating
$\epsilon$ and $p$ for a given value of $\mu$ :

\begin{equation}
\mu \equiv \mu_d = \mu_s = \mu_u + \mu_e ,
\end{equation}

\noindent by solving for $\mu_e$ from the condition that the total
electric charge of the system is zero.

There is an unphysical dependence of the equation of state on the
renormalization
point $\mu_o$, which, in principle, should not affect the
calculations of physical observables if the calculations are
performed to all orders in $\alpha_c$ \c{135,101}. In practice, the
calculations
are done perturbatively and, therefore, in order to minimize the
dependence on $\mu_o$ the renormalization point should be chosen to
be close to the natural energy scale, which could be either $\mu_o
\simeq B^{1/4}$ or the average kinetic energy of quarks in the bag, in
which case, $\mu_o \simeq$ 313 $MeV$. In the present study, our choice
of $\mu_o$ is dictated by the requirement that stable strange matter
occurs at zero temperature and pressure with a positive baryon
electric charge \c{150}. This leads to the following representative choice
of the parameter values:

\noindent Equation of state model 1 : $B = 56~ MeV~ fm^{-3}$;
m$_s$ = 150 $MeV$; $\mu_o$ = 150 $MeV$.

\noindent Equation of state model 2 : $B = 67$ $MeV$ fm$^{-3}$'
m$_s$ = 150 $MeV$; $\alpha_c$ = 0.

\noindent Model 2 corresponds to no quark interactions, but a non-zero mass
for the strange quark.

\ni In the limit, m$_s\rightarrow$ 0 and $\alpha_c\rightarrow$ 0, the equation
of state has the analytical form

\begin{equation}
p = \frac{1}{3} (\epsilon - 4B),
\end{equation}

\noindent where $\epsilon$ is the total energy density. Eq. (7.28) is
the MIT bag model. It is independent of the number of quark flavours.
\par
Numerical values of pressure $p$ and total mass-energy density $\rho
= \varepsilon/c^2$ for the quark matter equation of state models used here are
listed in Table 7.5. For the sake of comparison, we have included in
this table, the equation of state corresponding to non-interacting,
massless quarks
as given by the simple MIT bag model with $B = 56~MeV~fm^{-3}$.
Among these equation of state, the bag model is `stiffest' followed by
models 1 and
2. Equilibrium configurations of strange quark stars,
corresponding to the above equation of state, are
presented in Table 7.6, which lists the gravitational mass $M_G$, radius
$R$, the surface  red shift $z$, given by

\begin{table}
\cl{\bf TABLE 7.5}
\cl{ EQUATIONS OF STATE FOR DEGENERATE STRANGE
QUARK MATTER }
\vspace {0.2in}
\begin{center}
\begin{tabular}{cccc}
\hline
\multicolumn{1}{c}{$\rho (10^{14}~g~cm^{-3})$} &
\multicolumn{1}{c}{$P (10^{36}~dyn~cm^{-2})$} &
\multicolumn{1}{c}{} &
\multicolumn{1}{c}{} \\
\multicolumn{1}{c}{} &
\multicolumn{1}{c}{$model~1$} &
\multicolumn{1}{c}{$model~2$} &
\multicolumn{1}{c}{$MIT~bag~(B=56~MeV~fm^{-3})$}\\
\hline
6.0  & 4.44 & 2.23 & 6.01 \\
8.0 & 10.13 & 7.83 & 12.00 \\
10.0 & 15.88 & 13.49 & 17.99 \\
12.0 & 21.63 & 19.17 & 23.99 \\
14.0 & 27.41 & 24.88 & 29.98 \\
16.0 & 33.20 & 30.16 & 35.97 \\
18.0 & 39.00 & 36.36 & 41.96 \\
20.0 & 44.82 & 42.12 & 47.95 \\
22.0 & 50.64 & 47.89 & 53.95 \\
24.0 & 56.47 & 53.67 & 59.94 \\
26.0 & 62.30 & 59.46 & 65.93 \\
28.0 & 68.14 & 65.26 & 71.92 \\
30.0 & 73.98 & 71.06 & 77.91 \\
32.0 & 79.83 & 76.87 & 83.90 \\
36.0 & 91.53 & 88.51 & 95.89 \\
40.0 & 100.32 & 100.16 & 107.87 \\
50.0 & 132.58 & 129.36 & 137.83 \\
\hline
\end{tabular}
\end{center}
\end{table}
\begin{table}
\cl{\bf TABLE 7.6}
\vspace{0.2in}
\cl{ EQUILIBRIUM STRANGE QUARK STAR MODELS}
\vspace{0.2in}
\begin{center}
\begin{tabular}{cccccc}
\hline
\multicolumn{1}{c}{$Equation~ of~ state$} &
\multicolumn{1}{c}{$\rho_c (10^{14}~g~cm^{-3})$} &
\multicolumn{1}{c}{$M/M_{\odot}$} &
\multicolumn{1}{c}{$R(km)$} &
\multicolumn{1}{c}{$Surface~ red~ shift~(z)$} &
\multicolumn{1}{c}{$P_o(ms)$} \\
\hline
Model 1 & 24.0 & 1.958 & 10.55 & 0.487 & 0.488 \\
&20.0 & 1.967 & 10.78 & 0.472 & 0.503 \\
&16.0 & 1.951 & 11.02 & 0.448 & 0.522 \\
&12.0 & 1.864 & 11.22 & 0.401 & 0.548 \\
&8.0 & 1.521 & 11.02 & 0.299 & 0.591 \\
&6.0 & 0.997& 9.93  & 0.192 & 0.624 \\
&5.0 & 0.485 & 7.99 & 0.104 & 0.646 \\
\hline
Model 2 & 24.0 & 1.863 & 10.09 & 0.483 & 0.468 \\
&20.0 & 1.862 & 10.29 & 0.465 & 0.482 \\
&16.0 & 1.829 & 10.49 & 0.435 & 0.500 \\
&12.0 & 1.710 & 10.62 & 0.381 & 0.527 \\
&8.0 & 1.281 & 10.14 & 0.263 & 0.568 \\
&6.0 & 0.645 & 8.37 & 0.138 & 0.600 \\
&5.0 & 0.092 & 4.48 & 0.032 & 0.622 \\
\hline
MIT Bag & 24.0 & 2.021 & 10.81 & 0.493 & 0.500 \\
model &20.0& 2.033 & 11.04 & 0.480 & 0.514 \\
($B=56$ &16.0 & 2.023 & 11.29 & 0.450 & 0.533 \\
$MeV$ fm$^{-3}$)&12.0 & 1.947 & 11.52 & 0.410 & 0.558 \\
&8.0 & 1.635 & 11.41 & 0.310 & 0.604 \\
&6.0 & 1.150& 10.52 & 0.210 & 0.636 \\
&5.0 & 0.666 & 8.98 & 0.130 & 0.659 \\
\hline
\end{tabular}
\end{center}
\end{table}

\begin{equation}
z = (1 - 2 GM/c^2R)^{-1/2} - 1
\end{equation}

\ni and the period $P_o$ corresponding to fundamental
frequency $\Omega_o$ defined as \c{126} :

\begin{equation}
\Omega_o = \big(3 GM/4R^3\big)^{1/2}
\end{equation}

\noindent as functions of the central density $\rho_c$ of the star.

\ni We calculated the eigenvalue $\sigma$ by solving the Eq. (7.14) as
same way as discussed in the previous section.
\begin{figure}
\vskip 7.5in
\noindent{\bf Fig. 7.4}: Periods of radial pulsations as functions of the
gravitational mass. The top two and bottom left boxes correspond to
strange quark stars. The bottom right box is for stable neutron stars
corresponding to beta-stable neutron matter, model UV14 + UVII, ref.
\c{10}. The labels 1, 2, 3, 4, 5 correspond respectively to the
fundamental and the first four harmonics.
\end{figure}
\par
Results for the oscillations of quark stars corresponding to equation of state
models 1 and 2 are illustrated in Fig. 7.4. To compare,
we have included in Fig. 7.4 the results for quark stars corresponding
to (a) the simple MIT bag equation of state (non-interacting,
massless quarks and $B = 56~ MeV$ fm$^{-3}$) and (b) neutron stars
corresponding to a
recently given neutron matter equation of state \c{10}. The plots
in Fig. 7.4 are for
the oscillation time period (= 2$\pi/c\sigma$) versus the
gravitational mass $M_G$. The fundamental mode and the first four
harmonics are considered. The period is an increasing function of $M$,
the rate of increase being progressively less for higher oscillation
modes. The fundamental mode oscillation periods for quark stars are
found to have the following range of values:

\noindent MIT bag model  :  (0.14 - 0.32) $milliseconds$

\noindent Equation of state model 1  :  (0.10 - 0.27) $milliseconds$

\noindent Equation of state model 2  :  (0.06 - 0.30) $milliseconds$

\noindent For neutron stars (model UV14 + UVII, ref.\c{10}), we
find that the range of
periods for $l$ = 0 mode is (0.25 $-$ 0.4) $milliseconds$, which is
slightly less than the neutron star based on chiral sigma model
equation of state, presented in previous section. For higher
modes, the periods are $\leq$ 0.1 $milliseconds$, similar to the
case of quark stars but less than the chiral sigma model values.
So, the oscillation periods for neutron star based on chiral
sigma model equation of state is different from quark stars,
which is insignificant.
\par
Inclusion of strange quark mass and the quark interactions make the
equation of state a little `softer' as compared to the simple
MIT bag equation of state (Table 7.5). This is
reflected in the value of the maximum mass of the strange quark star
(Table 7.6).
For the pulsation of quark stars, this gives, for $l$=0 mode
eigenfrequencies, values as low as 0.06 $milliseconds$.
The main conclusion that emerges from our study,
therefore, is that use of realistic equation of state can be important in
deciding
the range of eigenfrequencies, at least for the fundamental mode of
radial pulsation. The
results presented here thus form an improved first step of calculations on
the lines presented by Haensel $et~al.$ \c{149}, whose numerical
conclusions are expected to get altered.
\par
Since we considered the vibrations of neutron stars and quark
stars, it is important to study the time scale for damping of
the vibrations. Regarding the time scale of damping of the
vibration, here we don't calculate exactly, but the damping time
is approximately same as Ref. \c{151}, considered by Madsen,
because, the oscillation time is taken to be $10^{-3}$$s$, which
is typical for the fundamental mode in our case. The discussion
by Madsen \c{151} was based on rather crude estimates.
\vfill
\newpage
\setcounter{equation}{0}
\chapter{Astrophysical application II : Cooling of neutron stars with
quark core}
\section{Introduction}
\hspace{0.3in} When a neutron star is formed in the collapse of
a stellar core,
it rapidly cools down by neutrino radiation. The interior
temperature drops to about less than $10^{10}~K$ within minutes and to about
$10^{9}~K$ within one year. Neutrino emission dominates the
subsequent cooling of the neutron star, until the interior temperature falls
to about $10^{8}~ K$, with a corresponding surface temperature of
about $10^{6}~ K$. Thus the photoemission also begin to
play an important role.
\par
The cooling curves (observed temperature as a function of time) depend
on a number of interesting aspects of the physics of neutron
star. It turns out
that the equation of state as well as the mass of the neutron
stars do not influence the
cooling in a sensitive manner, but the possible existence of a
superfluid state of the nucleons plays some role, and the existence
of a pion condensate or a quark phase in the central region would
have dramatic effects. For conventional cooling scenarios, neutrino
emission dominates the cooling for about $10^{5}$ years.
\par
The theoretical cooling models \c{151a,151b} have been refined \c{152,153} in
recent years by a number of authors. The search for thermal radiation
of pulsars
has so far in most cases has led only to upper bounds for the surface
temperature, still interesting comparisons between theory and
observations can be made.
\par
The so-called standard model of neutron star cooling is based
upon neutrino emission from the interior that is dominated by
the ${\it modified~ URCA~ process}$ \c{153a};

\begin{equation}
(n,p)+p+e^{-}\leftrightarrow (n,p)+n+\nu_{e},
\end{equation}

\begin{equation}
(n,p)+n\leftrightarrow (n,p)+p+e^{-}+\bar\nu_{e}.
\end{equation}

\ni The ${\it direct~~ URCA~~ process}$

\begin{equation}
n\rightarrow p+e^{-}+\bar\nu_{e},
\end{equation}
\begin{equation}
 p+e^{-}\rightarrow n+\nu_{e},
\end{equation}

\ni is not usually considered because it is strongly suppressed in
degenerate matter because of the requirement of energy and
momentum conservation. Since this
process has recently been revived \c{154}, we repeat the simple argument.
The fermions $n,~p,~e^{-}$ participating in the process have energies
lying within $T$ of the Fermi surface. By energy conservation, the
neutrino and antineutrino energies are then also $\sim T$. But the
Fermi momenta of the electrons and protons are small compared to
the neutron Fermi momentum and thus the processes (8.2) are strongly
suppressed by momentum conservation.
\par
This is no longer a case when an additional neutron, which can
absorb energy and momentum, takes part in the process, as in
(8.1), (8.2). A pion condensate would have the same effect as that of a
spectator neutron.
\par
It has recently been argued \c{154} that the proton concentration in a
neutron star might be so high that the momentum conservation, namely

\begin{equation}
p_{f}(p)+p_{f}(e) > p_{f}(n),
\end{equation}

\ni might be satisfied. For an $n,~p,~e$ mixture we have $n_{p}=n_{e}$ and
thus the condition becomes

\begin{equation}
n_{n}\le 8n_{p}.
\end{equation}

\ni The proton fraction $x=n_p/n$, where $n=(n_n+n_p)$ is the total baryon
density, is then given by

\begin{equation}
x\ge {1\over 9}\simeq 11.1\%
\end{equation}

\ni If the electron chemical potential exceeds the muon rest mass
$m_{\mu}=105.7~MeV$, muon will also be present in dense matter,
and this will increase the threshold proton concentration. If
$\mu_{e}\gg m_{\mu}$, the threshold proton concentration is
$\simeq 0.148$; for smaller values of $\mu_{e}$, the threshold
concentration lies between ${1\over 9}$ and 0.148. At densities
typical of the central regions of neutron stars, the calculated
proton concentration of matter is very sensitive to the choice of
physical model, and in reality it might exceed the threshold value
as Lattimer $et~al.$ \c{154} discuss. Estimates of proton fraction as a
function of baryon density for a number of different equation of
states indicate that, (a) the estimated proton concentrations depend
sensitively on the assumptions made about the microscopic
interactions, which are poorly known, and (b) it is quite
possible that the proton concentrations are large enough to allow the
direct Urca process to occur. As calculations by Wiringa, Fiks and
Fabrocini \c{10} demonstrate, the form of the three-body interaction,
especially its isospin dependence has a large influence on the
proton fraction.
\par
Let us now estimate the rate at which antineutrino energy emitted
per unit volume in the reaction given by (8.3). This may be done by using
Fermi's
``golden rule". Neglecting for the moment the effects of possible
superfluid of neutrons and superconductivity of protons, one finds

\begin{eqnarray}
\dot E_{\beta} = {2\pi\over{\hbar}}2\sum_{i}
G^{2}cos^{2}\theta_{C}(1+3g_{A}^{2})\nonu\\
\times n_{1}(1-n_{2})(1-n_{3})\times
\varepsilon_{4}\delta^{4}(p_1-p_2-p_3-p_4),
\end{eqnarray}

\ni where $n_{i}$ is the Fermi function and the subscript $i$=1 to 4 refer
to the neutron, proton, electron, and antineutrino respectively. The
$p_{i}$ are four-momenta, and $\varepsilon_{4}$ is the antineutrino
energy. The sum over states is to be performed only over possible
three momenta $p_{i}$ in unit volume, and prefactor 2 takes into
account the initial spin states of the neutron. The beta-decay matrix
element squared, after summing over spins of final particles and
averaging over angles, is obtained as $G^{2}cos^{2}\theta_{C}(1+3g_A^{2})$,
where $G=1.436\times 10^{-49}~ erg~ cm^{3}$ is the weak-coupling
constant, $\theta_{C}$ is the Cabibbo angle, and $g_A=-1.261$ being the
axial vector coupling constant. Final electron and proton states must
be vacant if the reaction is to occur, and this accounts for the
blocking factors $1-n_2$ and $1-n_3$. The electron-capture
process (8.4) gives the same energy loss rate as process (8.3), but in
neutrinos, and therefore the total luminosity per unit volume of the
Urca process is twice of Eq. (8.8). The integrals may be calculated
straightforwardly, since the neutrons, protons, and electrons are
very degenerate. One thus finds (Boltzmann's constant $k_B$ = 1)

\begin{equation}
\dot E_{Urca}={457\pi\over
10080}{{G^{2}cos^{2}\theta_{C}(1+3g_A^2)}\over{\hbar^{10}c^{5}}}m_n
m_p\mu_e (T)^{6}\Theta_t.
\end{equation}

\ni Here $\Theta_t$ is the threshold  factor $\Theta(p_e+p_p-p_n)$, which
is +1 if the argument exceeds 0, and is 0 otherwise.
\par
Particle interactions change this result in a number of ways. First,
the neutron and proton densities of states are determined by
effective mass rather than bare masses. Second, the effective
weak-interaction matrix elements can be modified by the medium. These
effects are expected to reduce the luminosity, but probably by less
than a factor of 10.
\par
The temperature dependence of the direct Urca emissivity may easily
be understood from phase-space considerations. The neutrino or
antineutrino momentum is $\sim ~ T$, and thus the phase space
available in final states might be a three-dimensional sphere of this
radius, whose volume is proportional to $( T )^3$. The
participating neutrinos, protons, and electrons are degenerate.
Therefore, for the reaction to occur, they must have energies that
lie within $\sim T$ of the energies at the Fermi surfaces, and
thus each degenerate particle contributes a factor $\sim  T$.
\par
Yet another possibility for the state of matter at high densities is
quark matter, in which quarks can move around essentially as free
particles, rather than being bound together as colour singlet
entities, such as nucleons and pions. Neutrino emission from such a
system was considered by Iwamoto \c{144}. The basic processes are
the quark analogue of the nucleon direct Urca processes (8.3)
and (8.4), $i.e.$,

\begin{equation}
d\rightarrow u + e^- + \bar\nu_e
\end{equation}
\ni and
\begin{equation}
u + e^{-}\rightarrow d + \nu_{e}.
\end{equation}

\ni The condition for beta equilibrium is

\begin{equation}
\mu_d=\mu_u+\mu_e.
\end{equation}

\ni If quarks and electrons are
treated as massless noninteracting particles, this condition is

\begin{equation}
p_{f}(d) = p_{f}(u)+p_{f}(e),
\end{equation}

\ni which is identical to the condition for it to be just possible to
conserve momentum for excitations near the respective Fermi surfaces.
At threshold the momenta of the $u$ quark, the $d$ quark, and the
electron must be collinear, but, as Iwamoto pointed out, the
weak-interaction matrix element for this case vanishes. However, if
quark-quark interactions are taken into account, the direct Urca
process for quarks and electrons that are
not collinear will be kinematically allowed. To illustrate this
effect, consider the case in which
interactions may be treated perturbatively. To first order in the
QCD coupling constant $\alpha$, the quark
chemical potentials are given by

\begin{equation}
\mu_{i}=[1+{8\over{3\pi}}\alpha]p_{f}(i),~ i=u,~d,
\end{equation}

\ni while the electron chemical potential is unchanged. Since $\alpha$ is
positive, it is easy to see that the conditions for beta
equilibrium [Eq. (8.12)] and momentum conservation may be satisfied
simultaneously. Angles characterizing deviations from collinearity
are typically of order $\alpha^{1/2}$. The calculation of the
neutrino and antineutrino emission rates proceeds in essentially the
same way as for the nucleon process, and the overall results for the
luminosity is

\begin{equation}
\dot E_{q}={914\over
315}{{G^{2}cos^{2}\theta_C}\over{\hbar^{10}c^{7}}}\alpha
p_{f}(d)p_{f}(u)\mu_e (T)^{6}.
\end{equation}

\ni This has a form similar to the nucleon Urca rate (8.9), but there are
some significant differences. First, there is a factor $\alpha$,
which reflects the fact mentioned above that the weak-interaction
matrix element vanishes for collinear relativistic particles, whereas
for non-relativistic nucleons the corresponding matrix element is
essentially independent of angle. The second difference is that the
quantities $p_{f}(u)$ and $p_{f}(d)$ take the place of the nucleon
masses. Third, the numerical coefficient is different because for
quarks the angular dependence of the matrix element is important.
However, since $p_{f}(u)$ and $p_{f}(d)$ are expected to be less than
$m_n$, and $\alpha$ is less than or of the order of unity, the
neutrino luminosity from quark matter is expected to be rather less
than the characteristic rate for the nucleon process. However, it is
important to note that for quark matter the electron fraction is
uncertain. For instance, if $u$, $d$, and $s$ quarks may be treated
as massless and free, the electron fraction vanishes identically.
Detailed estimates of the composition of quark matter for various
models are given by Duncan, Shapiro and Wasserman \c{145} and Alcock,
Farhi and Olinto \c{135}. So far, we have assumed the quark to be
massless. While this is a good approximation for $u$ and $d$ quarks,
but poor for $s$ quarks, which can participate in Urca process even
in the absence of strong interactions. However, detailed calculations
show that the energy-loss rate from processes in which $s$ quarks
participate is less than that from the processes where for $u$ and $d$
quarks are involved (Iwamoto \c{144}).

\section{Equilibrium neutrino emissivity of quark matter}
\hspace{0.3in} Iwamoto \c{144} has derived the formula for $\eps$
using apparently reasonable approximations and this formula has
been widely used \c{144,135,101,146} to calculate $\eps$ for
two and three flavour quark matter. According to his formula,
$\eps$ is proportional to baryon density $n_B$, strong coupling
constant $\alpha_c$ and sixth power of temperature $T$ for $d$
quark decay. For $s$ quark decay, $T$ dependence of $\eps$ is
same as that for $d$ decay. Furthermore,
his results imply that electron and quark masses have negligible
effect on $\eps$ and $s$ quark decay  (in
case of three flavour quark matter) which play a rather insignificant
role.
\par
In this section, we describe \c{154a} an exact numerical
calculation of $\eps$ and a comparison of our results
with the Iwamoto formula. Our results show that the Iwamoto
formula overestimates $\eps$ by orders of magnitude when
$p_{f}(u)+p_{f}(e)-p_{f}(d(s))$ is comparable with the temperature.
For reasonable values of $\alpha_{c}$ and baryon densities, this
quantity is much larger than the expected temperatures of neutron
stars ($\sim$ few 10ths of $MeV$) for two flavour quark matter, but is
comparable with temperature for three flavour quark matter.

\ni The neutrinos are emitted from the quark matter through reactions

\begin{eqnarray}
d \rightarrow u + e^{-} +\bar \nu_e \nonu \\
u+ e^- \rightarrow d+ \nu_e\nonu \\
s \rightarrow u + e^{-} +\bar \nu_e \nonu \\
u+ e^- \rightarrow s+ \nu_e.
\end{eqnarray}

\ni The equilibrium constitution of the quark matter is determined by
its baryon density $n_{B}$, charge neutrality conditions and weak
interactions given in Eq. (8.16). Thus, for two flavour quark matter,

\begin{eqnarray}
\mu_d = \mu_u + \mu_e   (\mu_{\nu_e}=\mu_{\bar\nu_e}=0)\nonu\\
 2 n_u - n_d -3 n_e = 0\nonu \\
n_B = ( n_u + n_d )/3
\end{eqnarray}

\ni and for three flavour quark matter

\begin{eqnarray}
\mu_d = \mu_u + \mu_e (\mu_{\nu_e}=\mu_{\bar\nu_e}=0) \nonu \\
\mu_d = \mu_s\nonu\\
 2n_u - n_d -n_s -3n_e = 0\nonu\\
n_B=(n_u+n_d+n_s)/3.
\end{eqnarray}

\ni The number density of species $i$ is  $n_i = g.p_f^3(i)/(6
\pi^2)$ with the degeneracy factor $g_{i}$ being two for electron and
six for quarks. For electrons
$\mu_{e}=\sqrt{p_{f}^2(e)+m_{e}^2}$ and for quarks we use \c{155}

\begin{equation}
\mu_q = [ {\eta \over x} + {8\alpha_c \over { 3 \pi}} ( 1- {3\over
{x \eta}} ln(x+\eta))]p_{f},
\end{equation}

\ni where $x\equiv p_{f}(q)/m_q$ and $\eta \equiv \sqrt{1+x^2}$,
$m_q$ being the quark mass. For massless quarks, Eq. (8.19) reduces to

\begin{equation}
\mu_q=(1+ {{8 \alpha_c}\over{3 \pi}}) p_{f}(q).
\end{equation}

\ni The neutrino emissivity $\eps$ for reactions involving
$d(s)$ quarks is calculated by using the reactions in Eq. (8.16). In
terms of the reaction rates of these equations, we get ( $\hbar~=~c~=~1$),

\begin{eqnarray}
\eps_{d(s)} =  A_{d(s)}\int d^3 p_{d(s)} d^3 p_u d^3 p_e d^3
p_{\nu} {( p_{d(s)}
. p_{\nu} ) ( p_u . p_e ) \over { E_u E_{d(s)} E_e }}\nonu \\
\times \delta^{4}(p_{d(s)}-p_u-p_e-p_{\nu}) n(\vec p_{d(s)}) [1-n(\vec
p_u)][1-n(\vec p_e)],
\end{eqnarray}

\ni where $p_i=(E_i,~\vec p_i)$ are the four momenta of the particles,
$n(\vec p_i)~=~ {1\over{ e^{\beta (E_i - \mu_i)} +1 }}$ are the Fermi
distribution functions and

\begin{eqnarray}
A_d = {24 G^2 \cos^2\theta_c\over{(2 \pi )^8}}\\
A_s = {24 G^2 \sin^2\theta_c\over{(2 \pi )^8}}.
\end{eqnarray}

\ni For degenerate particles, ($\beta p_{f}(i)\gg 1$), Iwamoto has
evaluated the integrals in Eq. (8.21) using certain reasonable
approximations and obtained the simple expressions for $\eps_d$
and $\eps_s$ as given below \c{144},

\begin{eqnarray}
\eps_d = {{914\over 315} {G^2 \cos^2\theta_c\alpha_c p_{f}(d) p_{f}(u)
p_{f}(e) T^6}} \nonu \\
\eps_s = {{457 \pi \over 840} {G^2 \sin^2\theta_c\alpha_c \mu_{s} p_{f}(u)
p_{f}(e) T^6}}.
\end{eqnarray}

\ni The approximations involved in obtaining these formulas are \\

\begin{enumerate}
\item neglect of neutrino momentum in momentum conservation,
\item replacing the matrix elements by some angle averaged  value, \\
and
\item decoupling  momentum and angle integrals.
\end{enumerate}
\par
The expressions for neutrino emissivity as obtained by Iwamoto have
been used widely. The temperature dependence of emissivity as
obtained by Iwamoto has a physical explanation. Each
degenerate fermion gives one power of $T$ from the phase space integral
( $d^3 p_i \rightarrow {p_{f}(i)}^2 dE_i\propto T$ ). Thus one gets
$T^3$ from quarks and electrons. Phase space integral for the
neutrino gives $d^3 p_{\nu} \propto (E_{\nu}^2) dE_{\nu} \propto
T^3$. Energy conserving $\delta-$function gives one $T^{-1}$ which is
cancelled by one $E_\nu~\propto~T$ factor coming from matrix element.
So finally, one gets $\eps \propto T^6$. This argument, however,
ignores the fact that
$\Delta p_{d}~(\Delta p_{s})~=~p_{f}(u)+p_{f}(e)-p_{f}(d)~(p_{f}(s))$,
which
is related to the angle between $\vec p_d$, $\vec p_u$ and $\vec
p_e$ could be small and comparable to $T$. We shall demonstrate below
that precisely in this region that the Iwamoto formula fails.
\par
Before discussing the causes of the shortcoming of Iwamoto formula, let us
first compare our results with the Iwamoto formula
and try to find out the specific cases where the deviation is more
pronounced. In Figs.8.1-8.3, we have plotted $\eps$ vs $T$
for two-flavour $d$ decay, three-flavour $d$ decay and $s$ decay respectively.
For two-flavour $d$ decay our results ($\eps_d$) are in good
agreement with  the emissivity calculated using Iwamoto formula
($\eps_{dI}$). In Fig. 8.1 curves (a) and (b) are $\eps_{dI}$
and $\eps_{d}$ respectively, for $\alpha_c$ = 0.1 and $n_B$ =
0.4. (c) and (d) corresponds to the same but for $\alpha_c$= 0.1 and
$n_B$= 1.4. It is evident from the figure that agreement of Iwamoto
results with our calculation is better for higher densities and lower
temperatures. Also $\eps_d$ is consistently smaller than
$\eps_{dI}$, the Iwamoto result, in the range of temperatures
considered. Corresponding Fermi momenta of quarks and electrons are
given in Table 8.1. It is to be noted that all the momenta are much
larger than the temperature.

\begin{table}
\cl{\bf TABLE 8.1}
\vspace{0.2in}
\cl{BARYON NUMBER DENSITY $n_B$, FERMI MOMENTA OF $u$-QUARK
$p_{f}(u)$,}
\cl{ $d$-QUARK $p_{f}(d)$, AND ELECTRON $p_{f}(e)$ FOR DIFFERENT $\alpha_c$,}
\cl{ WHERE $\Delta p_d=p_{f}(u)+p_{f}(e)-p_{f}(d)$.}
\vspace {0.2in}
\begin{center}
\begin{tabular}{cccccc}
\hline
\multicolumn{1}{c}{$\alpha_c$} &
\multicolumn{1}{c}{$n_B$} &
\multicolumn{1}{c}{$p_{f}(u)$} &
\multicolumn{1}{c}{$p_{f}(d)$} &
\multicolumn{1}{c}{$p_{f}(e)$}&
\multicolumn{1}{c}{$\Delta p_d$}\\
\multicolumn{1}{c}{} &
\multicolumn{1}{c}{($fm^{-3}$)} &
\multicolumn{1}{c}{($MeV$)} &
\multicolumn{1}{c}{($MeV$)} &
\multicolumn{1}{c}{($MeV$)} &
\multicolumn{1}{c}{($MeV$)} \\
\hline
        &0.60&357.80&449.20&99.15&7.75\\
     0.1&1.00&424.22&532.52&117.56&9.20\\
        &1.40&474.57&595.79&131.51&10.29\\
\hline
        &0.60&357.71&449.25&95.43&3.89\\
     .05&1.00&424.11&532.65&113.15&4.61\\
        &1.40&474.45&595.87&126.57&5.15\\
\hline
\end{tabular}
\end{center}
\end{table}

\begin{figure}
\vskip 7.5in
\noindent {\bf Fig. 8.1}: Two flavour $d$-decay for $\alpha_c=0.1$;
(a) Iwamoto  results for $n_B=0.4 fm^{-3}$, (b) Our
results for $n_B=0.4 fm^{-3}$ ($\Delta p_d=6.78$), (c) Iwamoto
results for $n_B=1.4 fm^{-3}$, (d) Our results for $n_B=1.4
fm^{-3}$ ($\Delta p_d=10.29$).
\end{figure}
\begin{figure}
\vskip 7.5in
\noindent {\bf Fig. 8.2}: Three flavour $d$-decay for
$\alpha_c=0.1$ and $s$ quark mass is 150 $MeV$; (a) Our
results for $n_B=1.4 fm^{-3}$, (b) Iwamoto results for $n_B=1.4
fm^{-3}$ ($\Delta p_d=0.067$), (c) Our results for $n_B=0.4 fm^{-3}$, (d)
Iwamoto results for $n_B=0.4 fm^{-3}$ ($\Delta p_d=0.39$).
\end{figure}
\begin{figure}
\vskip 7.5in
\noindent {\bf Fig. 8.3}: Three flavour $s$-decay for
$\alpha_c=0.1$ and $s$ quark mass is 150 $MeV$; (a) Our
results for $n_B=1.4 fm^{-3}$, (b) Iwamoto results for $n_B=1.4
fm^{-3}$ ($\Delta p_s=1.613$), (c) Our results for $n_B=0.4 fm^{-3}$, (d)
Iwamoto results for $n_B=0.4 fm^{-3}$ ($\Delta p_d=9.719$).
\end{figure}
\par
Fig. 8.2 shows the $\eps_d$ for three-flavour quark matter. It shows that
$\eps_{dI}$ is 2 -3 orders of magnitude higher compared to our
results. Here contrary to the two flavour case, the difference
becomes more pronounced at higher densities. Fig. 8.3 shows the
variation of $\eps_s$ with temperature. Here again, it is clear
that $\eps_s$ is quite different from $\eps_{sI}$ but this
difference is less compared to that between $\eps_d$ and
$\eps_{dI}$. For all the cases, the difference between our results
and those using Iwamoto formula increases at higher temperatures.
The Fermi momenta of quarks and electron for three- flavour case are
given in Table 8.2. The
study of all the figures and tables above reveals that the cases
where Iwamoto formula agrees reasonably well with our results,
$\Delta p_d$ (or $\Delta p_s$ ) is much larger than the temperature.
On the other hand, when this difference is smaller or comparable with
the temperature, the Iwamoto formula overestimates the exact result
by order of magnitude. In addition to these, Fig. 8.3 also shows that
our results are about a factor of 2.5 lower than the Iwamoto results
even at lower temperatures. We have found that this difference comes
from the approximation involved in the calculation of matrix element.

\begin{table}
\cl{\bf TABLE 8.2}
\vspace{0.2in}
\cl{ BARYON NUMBER DENSITY $n_B$, FERMI MOMENTA OF $u$-QUARK
$p_{f}(u)$, $d$-QUARK}
\cl{ $p_{f}(d)$, $s$-QUARK $p_{f}(s)$ AND ELECTRON $p_{f}(e)$ FOR
DIFFERENT $m_s$ AND}
\cl{ DIFFERENT $\alpha_c$, WHERE $\Delta p_d=p_{f}(u)+p_{f}(e)-p_{f}(d)$}
\cl{ AND $\Delta p_s=p_{f}(u)+p_{f}(e)-p_{f}(d)$ }
\vspace {0.2in}
\begin{center}
\begin{tabular}{ccccccccc}
\hline
\multicolumn{1}{c}{$m_s$} &
\multicolumn{1}{c}{$\alpha_c$} &
\multicolumn{1}{c}{$n_B$} &
\multicolumn{1}{c}{$p_{f}(u)$} &
\multicolumn{1}{c}{$p_{f}(d)$} &
\multicolumn{1}{c}{$p_{f}(s)$} &
\multicolumn{1}{c}{$p_{f}(e)$}&
\multicolumn{1}{c}{$\Delta p_d$}&
\multicolumn{1}{c}{$\Delta p_s$} \\
\multicolumn{1}{c}{($MeV$)} &
\multicolumn{1}{c}{} &
\multicolumn{1}{c}{($fm^{-3}$)} &
\multicolumn{1}{c}{($MeV$)} &
\multicolumn{1}{c}{($MeV$)} &
\multicolumn{1}{c}{($MeV$)} &
\multicolumn{1}{c}{($MeV$)}&
\multicolumn{1}{c}{($MeV$)}&
\multicolumn{1}{c}{($MeV$)} \\
\hline
     &   &0.60&356.99&360.06&353.86&3.33&0.26&6.46\\
     &0.1&1.00&423.26&424.81&421.69&1.69&0.14&3.26\\
150.0&   &1.40&473.49&474.27&472.72&0.84&0.06&1.61\\
\cline{2-9}
     &    &0.60&356.99&365.89&347.62&9.28&0.38&18.65\\
     &0.05&1.00&423.26&430.29&415.98&7.33&0.30&14.61\\
     &    &1.40&473.49&479.49&467.34&6.25&0.25&12.40\\
\hline
     &   &0.60&356.99&365.93&347.58&9.70&0.76&19.11\\
     &0.1&1.00&423.26&429.10&417.25&6.34&0.50&12.35\\
200.0&   &1.40&473.49&477.66&469.25&4.52&0.35&8.76\\
\cline{2-9}
     &    &0.60&356.99&374.26&337.85&18.00&0.73&37.14\\
     &0.05&1.00&423.26&437.14&408.39&14.47&0.59&29.34\\
     &   &1.40&473.49&485.45&460.90&12.46&0.50&25.05\\
\hline
\end{tabular}
\end{center}
\end{table}
\par
Furthermore, Table 8.2. shows that for three flavour case electron
chemical potential (which is same as $p_{f}(e)$ for massless
electrons ) becomes small ($<~$1~$MeV$  ) for some values of $\alpha_{c}$,
$n_{B}$ and $m_{s}$. In these cases, electrons are no longer degenerate.
Clearly, for such cases the Iwamoto formula is not applicable. This
point is missed in earlier calculations.
\par
Our results are expected to have important implications for
neutrino emissivity and
quark star cooling rates because all the earlier calculations have
used the Iwamoto formula and predicted large quark star cooling rates
in comparison with the neutron star cooling rates for temperatures
less than 1 $MeV$. Our results show that particularly, for three-flavour
quark matter, the calculated emissivity is at least two orders of
magnitude smaller than the one given by Iwamoto formula and
therefore, the three-flavour quark star cooling rates are that much
smaller. Hence, it is necessary to understand why Iwamoto formula fails.
\par
To investigate the shortcoming of the Iwamoto formula, we consider the integral

\begin{eqnarray}
I = \int \frac {d^3 p_d d^3 p_u d^3 p_e d^3p_\nu}
{\eps_d\eps_u\eps_e}\nonu\\
\times \delta^{4}(p_{d(s)}-p_u-p_e-p_{\nu}) n(\vec p_{d(s)})
[1-n(\vec p_u)][1-n(\vec p_e)].
\end{eqnarray}

\ni Here, we have replaced the neutrino emission rate by
unity and therefore $I$ is essentially the phase space integral.
Following the reasoning of Iwamoto, this integral should be
proportional to $T^5$. Choosing the coordinate axes such that $\vec
p_d$ is  along $z$-axis and $\vec p_u$ is in $x-z$ plane and using the
three-momentum $\delta-$function to perform electron and $u$-quark angle
integrations, we get

\begin{eqnarray}
I = 8\pi^2\int \frac {p_d^2d p_d p_u^2dp_u p_e^2dp_e d^3p_\nu}
{\eps_d\eps_u\eps_e} \nonu\\
\times [\frac {\sqrt{1-x_u^2}}{p_ep_u(\sqrt{1-x_u^2}(p_d-p_\nu x_\nu)+
p_\nu x_\nu\sqrt{1.-x_\nu^2}\cos\phi_\nu} ] \nonu \\
\times \delta(\eps_{d}-\eps_u-\eps_e-\eps_\nu) n(\vec p_d) [1-n(\vec
p_u)][1-n(\vec p_e)],
\end{eqnarray}

\ni where $x_\nu~=~\cos\th_\nu$ and $x_u~=~\cos\th_u$ is determined by
solving

\begin{eqnarray}
p_ux_u~&=&~p_d-p_\nu x_\nu~-~ [p_e^2-p_u^2(1-x_u^2)-p_\nu^2(1-x_\nu^2)
\nonu \\
{}~&~&~~~-2p_up_\nu\sqrt{(1-x_u^2)(1-x_\nu^2)}\cos\phi_\nu]^{1/2}.
\end{eqnarray}

\ni The integral in Eq. (8.26) above is restricted to the momenta
$|p_i~-~p_f(i)|~ $ few times $T$ due to Fermi distribution
functions and the energy $\delta-$function.
Now, if we neglect the neutrino momentum in the $\delta-$functions,
we get, $x_u~=~(p_d^2+p_u^2-p_e^2)/2p_dp_u$ and the factor in the
square brackets of Eq. (8.26) becomes $1/p_dp_up_e$.

\ni Two points should be noted at this stage.

\begin{enumerate}
\item  Generally, $x_u$ is close to unity, so that $1-x_u^2$
is small. But, if $\Delta p_d$ is of the order of $T$,
$\sqrt{1-x_{u}^2} p_d$ can be comparable with $T$ and $p_{\nu}$ and
therefore $p_{\nu}$ cannot be neglected in the momentum
$\delta-$functions. Particularly, the denominator in the square
bracket of Eq. (8.26) cannot be approximated by $p_ep_up_d\sqrt{1-x_u^2}$.
Thus, if $p_d\sqrt{1-x_u^2}~<~p_\nu$, one would get a power of $T$ from
the denominator and $I$ will not be proportional to $T^5$.
\item Secondly, the momenta may differ from the
corresponding Fermi momenta by few times $T$ in the integral. When
$\Delta p_d~\sim~T$,  there are
regions in $p_dp_up_e-$space where $x_u~>~1$ and the rest of the
integrand is not small. Clearly, these regions must be excluded from
the integration as these values of $x_{u}$ are unphysical. If one does
not put this restriction, as is done when one factorises
angle and momentum integrals, the phase space integral will be
overestimated (and wrong) when $\Delta p_d~\sim~T$.
\end{enumerate}
\par
The above discussion clearly shows why the integral in Eq. (8.26) should
not be proportional to $T^5$ when $\Delta p_d~\sim~T$. In order to
demonstrate this point, we have calculated the integral in Eq. (8.26)
numerically and compared with the approximation where the neutrino
momenta are neglected and the restriction imposed by $x_u$ condition
is not imposed. The calculation is done for $\alpha_c=0.1$ and for
two-flavour case. The results are shown in Fig. 8.4. In this figure, we
also show the result for a case where the electron mass is taken to
be 25 $MeV$. This is of course unphysical, but by adjusting the electron
mass we can reduce $\Delta p_d$. The figure clearly shows that the
approximate value of $I$ is proportional to $T^5$ where as the exact
integral is smaller than the approximate one at large $T$. Further
more, for 25 $MeV$ electron mass, the departure from $T^5$ sets in at
smaller value of the temperature. This clearly shows that the
departure is dependent on the value of $\Delta p_d$. Here we would
like to mention that for some values of $\alpha_c$  and $m_{s}$
, $p_{f}(e)$ is small and is of the order of $T$. This
implies that electrons are no longer degenerate and deviation from
the Iwamoto result is most pronounced.

\begin{figure}
\vskip 7.5in
\noindent {\bf Fig. 8.4}: Two flavour phase space integrals for
$\alpha_c=0.1$ (a) Without restriction on $\cos\theta_u$ for
both electron mass $m_e$=0.0 and 25 $MeV$, (b) Exact integral for
$m_e$=0.0, (c) Exact integral for $m_e$=25 $MeV$.
\end{figure}
\begin{figure}
\vskip 7.5in
\noindent {\bf Fig. 8.5}: $\epsilon_{d(s)I} \over {\epsilon_{d(s)}}$
is plotted against $x$ where $x \ = \ {T\over {\Delta p_{d(s)}}}$.
The fitted function is $f(x) \ = \ 1~+~ax~+~bx^2~+~cx^3$
where $a=-2.5$, $b=100.$ and $c=30.$
\end{figure}
\par
In Eq. (8.26), we have dropped the matrix element of the weak interaction
in the emissivity calculation (Eq. (8.21)). So, the discussion of
preceeding paragraphs apply to the emissivity calculation as well.
Therefore, it is now clear why the Iwamoto formula
fails, when $\Delta p_d$ (or $\Delta p_s$ in case of weak interactions
involving strange quarks) is close to the temperature of the quark
matter.
\par
Similar approximations have been used by other authors \c{156,157}. In
Ref. \c{156}, Burrows has calculated the neutrino emissivity for
non-interacting quark matter from reverse beta decay. The exponent of
$T$ is 7 instead of 6 (Iwamoto), because of the
partial restriction of the electron's phase space. Duncan
$et~al.$ \c{157} computed the emissivity from reverse and direct beta
decay for both interacting and non-interacting quark matter. They
have reproduced the results of Burrows ($T^7$) \c{156} for
non-interacting reverse beta decay and Iwamoto ($T^6$) \c{144} for
interacting quark matter. In both these works, the effect of the
finite neutrino momentum is included but rest of the calculation follows the
approximation scheme of Iwamoto. Hence, their temperature dependence
of emissivity formula is different from ours.
Here, we would like to note that the departure from $T^6$ dependence
of the emissivity essentially arises from the careful phase space
integration. Since, similar approximation scheme is used to obtain
the neutrino emissivity of neutron matter, it is possible that the emissivity
calculated for neutron matter may also be overestimated when $x$ is
large.
\par
Since the departure from the Iwamoto formula arises from the fact
that $T / \Delta p_{d}$ (or $T / \Delta p_{s}$ ) is not small, it may be
possible to fit the numerically calculated $\eps$ with a
function of the form $\eps_{I}/f(x)$, where $x \ = \ T / \Delta p_{d}$
($T / \Delta p_{s}$ for strange sector). The function $f(x)$ should be
such that for small values of $x$ it should approach unity. Choosing
$f(x) \ = \ 1 \ + \ a x \ + \ bx^2 \ + \ cx^3$, we have
fitted the
calculated $\eps$ for a number of values of $n_{B}$, $\alpha_{c}$
and $ m_{s}$ for both $s$ and $d$ decay and obtained the values of
$a$, $b$ and $c$. The quality of fit is shown in Fig. 8.5 (since,
for $s$- decay, as mentioned above, there is difference of a factor
of 2.5 in Iwamoto and our results even at lower temperatures, the
data points for $s$- decay, in Fig. 8.5., have been scaled
accordingly). The values of $a$, $b$ and $c$ are $-2.5$, $100.$ and
$30.$ respectively. It is clear from Fig. 8.5 that for $x \rightarrow
0$, Iwamoto results approache to our values. Hence, our fitting is
valid for any values of $x$.
\vfill
\newpage
\section{Non-equilibrium neutrino emissivity of quark matter}
\hspace{0.3in}
The neutrino emission is a dominating mechanism of cooling
of quark or neutron stars with the internal temperature
exceeding $\simeq 10^{8} K$. The quark stars cool much faster
than the neutron stars.
The neutrino emissivity formula \c{144,154a} of the quark stars
are obtained under the assumption that the system is in
equilibrium with respect to weak interactions. However, as long
as matter is strongly degenerate, the number density of
neutrinos is small and their presence in the matter can
be neglected. Thus, the beta equilibrium (chemical equilibrium) stands
as a good approximation, which implies the equality of anti-neutrino and
neutrino emissivity. Of course, one can work even though the
chemical equilibrium is not satisfied. Radial pulsations of
quark stars have periods $\sim$ $milliseconds$, much shorter than the
time scale of the beta reaction at $T < 10^{10} K~\sim 1~MeV$. Local
compression
as well as the rarefaction are very much important for non-equilibrium beta
reactions and depend on the difference between the chemical
potentials of $u$, $d(s)$ quarks and e electron, $i.e.$,
$\delta\mu=\mu_{d(s)}-\mu_{u}-\mu_{e}$. The non-equilibrium beta
reactions, induced by the radial pulsations of neutron and quark stars,
were studied by a number of authors \c{126,149}. Recently, Madsen \c{151}
and Sawyer \c{147} calculated the bulk viscosity and damping rates
by invoking the non-equilibrium  condition in the quark star
matter. However, later, Haensel \c{159} calculated the neutrino
emissivity of non-equilibrium neutron star matter, where he
found that the emissivity depends on the non-equilibrium
conditions such as the difference in the chemical potential of
neutron, proton and electron
($\delta\mu^{'}=\mu_{n}-\mu_{p}-\mu_{e}$). Moreover, he showed that with
increase of $\delta\mu^{'}(-\delta\mu{'})$ the emissivity
increases (decreases). So, it is
interesting to see the  non-equilibrium effects in the quark star matter.
\par
In this section, we consider the rapid compression of the liquid interior
in quark stars, which has two/three component quark matter. Such
a situation is expected to occur during the gravitational collapse of
the neutron stars or the quark stars in black hole \c{160}, which
would take place when the mass of an accreting neutron star exceeds
the maximum allowable mass for the equilibrium configurations.
Thus, a significant deviation from the chemical equilibrium is to be expected
because of the shrinking of the stellar radius, which implies the monotonic
increase of the average density in collapsing stars. Here we
concentrate \c{161} on the characteristic features of the beta
non-equilibrium neutrino spectra and compared with that of the beta
equilibrium neutrino spectra.
\par
Let us consider the neutrino emissivity of quark matter having
two/three flavour degrees of freedom. Each of the constituents
separately is a  Fermi liquid in thermodynamic equilibrium.

\ni For the degenerate two flavour quark matter, the simplest neutrino
processes are the direct beta decay reactions

\begin{eqnarray}
 direct~\beta_{-} : d \rightarrow u + e^{-} +\bar \nu_e \nonumber \\
direct~\beta_{+} : u+ e^- \rightarrow d+ \nu_e.
\end{eqnarray}

\ni The distinction between  $\beta_{-}$ and $\beta_{+}$  processes will
be discussed later on.

\ni Moreover, the charge neutrality process of the two flavour quark
matter is

\begin{equation}
 2 n_u - n_d -3 n_e = 0
\end{equation}

\ni and the baryon density is defined as $n_B = ( n_u + n_d )/3 $ .
Since the matter is strongly degenerate, we write the Fermi momentum
${p_{F}}_{i}=(6 \pi^2 n_{i}/g)^{1/3}$ by neglecting the
thermal corrections,
$n_{i}$ being the number density of $i$-th particle and the
degeneracy factor $g$ is 6 for quark and 2 for electron.
However, the non-equilibrium condition for the above reaction is

\begin{equation}
\mu_d - \mu_u - \mu_e = \delta\mu   (\mu_{\nu_e}=\mu_{\bar\nu_e}=0).
\end{equation}

\ni Here, we approximate the positive $\delta\mu$ for the direct $\beta_{-}$
reaction (anti-neutrino emission) and that of the negative $\delta\mu$ for
the direct $\beta_{+}$ reaction (neutrino emission).

\ni Similarly, for the three flavour degenerate quark matter, the simplest
neutrino process are the direct beta decay reaction, which occurs
through $d$ and $s$ quarks. In addition to Eq. (8.28), one has also
the following reactions :

\begin{eqnarray}
direct~\beta_{-} : s \rightarrow u + e^- + \bar\nu_e \nonumber \\
direct~\beta_{+} : u + e^- \rightarrow s + \nu_e
\end{eqnarray}

\ni The equations for the charge neutrality and the baryon
density are respectively

\begin{equation}
 2n_u - n_d -n_s -3n_e = 0
\end{equation}

\ni and

\begin{equation}
n_B=(n_u+n_d+n_s)/3,
\end{equation}

\ni and due to the non-equilibrium condition, one can write

\begin{eqnarray}
\mu_d - \mu_s = \delta\mu.
\end{eqnarray}

\ni Moreover, here we again approximate the positive $\delta\mu$
for the direct
$\beta_{-}$ reaction (anti-neutrino emission) and the negative $\delta\mu$
for that of the direct  $\beta_{+}$ reaction (neutrino emission).
\par
Thus, in these approximations for both the two and the
three flavour quark  matter, the neutrino emissivity is almost
same, $i.e.$, $\varepsilon_{\bar\nu_{e}}(-\delta\mu)\simeq
\varepsilon_{\nu_{e}}  (\delta\mu)$. In the chemical equilibrium,
$\delta\mu = 0 $, which implies $\varepsilon_{\bar\nu_{e}} \simeq
\varepsilon_{\nu_{e}}$.
\par
The energy momentum relation of $s$ quark is approximated by
Eq. (8.19) and that of $u$ and $d$ quarks by Eq. (8.20), $i.e.$, $\mu$
is replaced by $E$ and $p_F$ is replaced by $p$, in above Eqs. (8.19-8.20).
This is reasonable since only the energies and the momenta close to
the Fermi surface will contribute to the matrix element.
The neutrino emissivity $\varepsilon_{d(s)}$ for $d(s)$ decay \c{144} is given
by Eq. (8.21). The momentum delta function in  Eq. (8.21) is
used to integrate over the
neutrino momentum $p_\nu$. For the massless electrons, $E_e = p_e$ and the
energy delta function is used to perform the integral over $p_e$. This gives

\begin{equation}
p_e= {{(E_{d(s)}-E_u)^2- (p_{d(s)}-p_u)^2- 2 p_{d(s)} p_u
(1-cos\theta_{u})}\over{ (2 p_u cos\theta_{ue}+ 2 p_{d(s)}
cos\theta_e+ 2 E_{d(s)}- 2 E_u)}},
\end{equation}

\ni where $\theta_{u}$, $\theta_{e}$ and $\theta_{ue}$ are the
angles between $d$ and $u$, $d$ and $e$ and $u$ and $e$ respectively.
\par
The chemical potentials and Fermi momenta thus obtained by
employing the above non-equilibrium and charge neutrality conditions
for two and three flavour matter  are substituted in neutrino emissivity
expression [Eq. 8.21] which subsequently yield

\begin{eqnarray}
\varepsilon_{d(s)} = A_{d(s)}\int d^3 p_{d(s)} d^3 p_u p_{e}^{2} d\Omega_{e}
 {( p_{d(s)}. p_{\nu} ) ( p_u . p_e ) \over { E_u E_{d(s)} E_e
}} \nonu \\
\times n(\vec p_{d(s)}) [1-n(\vec p_u)][1-n(\vec p_e)].
\end{eqnarray}
\par
In the present case, the above emissivity has been evaluated
numerically. The integral is five
dimensional as all the angles are measured with respect
to $d(s)$ quark. Throughout the calculation, we have
taken $u,~d,~e,$ and $\nu$ masses to be zero, $s$ quark mass $m_s$
and strong coupling constant $\alpha_c$ to be 150 $MeV$ and
0.1 respectively.
\par
Recently \c{154a}, the neutrino emissivity of the degenerate quark matter
was calculated exactly with the chemical equilibrium
($\delta\mu=0.0$), which has been discussed in the previous
section. It has been found that the neutrino emissivity
results are in qualitative agreement with that of Iwamoto \c{144}
for the two flavour quark matter, whereas for the three flavour quark
matter Iwamoto's result overestimates the
numerical values by nearly 2 orders of magnitude or more for $d$ decay
and agrees with that of the $s$-decay results within a factor of 3-4.
Also, it was pointed out that the dependence of the temperature and
the density on the neutrino emissivity
is quite different from Iwamoto's result and is sensitive to the $s$ quark
mass.
\par
Here, we use the same method \c{154a} to calculate the neutrino emissivity by
employing the non-equilibrium condition. For the two values of
temperatures as well as baryon densities and different values of
$\delta\mu$, the numerical values of
the neutrino emissivity for two and three flavour quark matter are quoted
in Table 8.3 and 8.4 respectively. The variation of the corresponding Fermi
momenta with respect to the increase and decrease of $\delta\mu$ are
presented in Table 8.5 and 8.6. We found here that the dependence of
emissivity on $n_{B}$ and
$T$ is rather different from that of the equilibrium condition ($\delta\mu=0$).

\begin{table}
\cl{\bf TABLE 8.3}
\vspace{0.2in}
\cl{ EMISSIVITY FOR TWO FLAVOUR QUARK MATTER. HERE
$\epsilon_{\bar\nu}^{d}(\delta\mu)$ AND}
\cl {$\epsilon_{\bar\nu}^{d}(-\delta\mu)$ ARE THE EMISSIVITIES FOR
$d$ DECAY.}
\vspace {0.2in}
\begin{center}
\begin{tabular}{ccccc}
\hline
\multicolumn{1}{c}{T} &
\multicolumn{1}{c}{$n_B$} &
\multicolumn{1}{c}{$\epsilon_{\bar\nu}^{d}(-\delta\mu)$} &
\multicolumn{1}{c}{$\epsilon_{\bar\nu}^{d}(\delta\mu)$} &
\multicolumn{1}{c}{$\delta\mu$}\\
\hline
\multicolumn{1}{c}{($MeV$)} &
\multicolumn{1}{c}{$(fm^{-3})$} &
\multicolumn{1}{c}{$(erg/cm^{3}/s)$} &
\multicolumn{1}{c}{$(erg/cm^{3}/s)$} &
\multicolumn{1}{c}{$fm^{-1}$} \\
\hline
   &0.8&7.64$\times 10^{+27}$&1.20$\times 10^{+31}$&0.005 \\
   &1.0&9.59$\times 10^{+27}$&1.54$\times 10^{+31}$&     \\
\cline{2-5}
    &0.8&7.22$\times 10^{+26}$&4.50$\times 10^{+31}$&0.0075 \\
    &1.0&9.02$\times 10^{+26}$&5.80$\times 10^{+31}$&     \\
\cline{2-5}
0.4 &0.8&3.75$\times 10^{+25}$&1.36$\times 10^{+32}$&0.01 \\
    &1.0&4.52$\times 10^{+25}$&1.79$\times 10^{+32}$&     \\
\cline{2-5}
    &0.8&4.50$\times 10^{+29}$&4.50$\times 10^{+29}$&0.00 \\
    &1.0&5.70$\times 10^{+29}$&5.70$\times 10^{+29}$&     \\
\hline
   &0.8&6.67$\times 10^{+28}$&2.33$\times 10^{+31}$&0.005\\
   &1.0&8.41$\times 10^{+28}$&3.02$\times 10^{+31}$&     \\
\cline{2-5}
    &0.8&1.12$\times 10^{+28}$&7.10$\times 10^{+31}$&0.0075\\
    &1.0&1.42$\times 10^{+28}$&9.29$\times 10^{+31}$&     \\
\cline{2-5}
0.5 &0.8&1.58$\times 10^{+27}$&1.86$\times 10^{+32}$&0.01 \\
    &1.0&1.97$\times 10^{+27}$&2.49$\times 10^{+32}$&     \\
\cline{2-5}
    &0.8&1.61$\times 10^{+30}$&1.61$\times 10^{+30}$&0.00 \\
    &1.0&2.05$\times 10^{+30}$&2.05$\times 10^{+30}$&     \\
\hline
\end{tabular}
\end{center}
\end{table}
\begin{table}
\cl{\bf TABLE 8.4}
\vspace{0.2in}
\cl{ EMISSIVITY FOR THREE FLAVOUR QUARK MATTER. HERE
$\epsilon_{\bar\nu}^{d}(\delta\mu)$ AND
$\epsilon_{\bar\nu}^{d}(-\delta\mu)$}
\cl { ARE THE EMISSIVITIES  FOR $d$ DECAY AND
$\epsilon_{\bar\nu}^{s}(\delta\mu)$ AND $\epsilon_{\bar\nu}^{s}(-\delta\mu)$}
\cl{ ARE THE EMISSIVITIES FOR $s$ DECAY.}
\vspace{0.2in}
\begin{center}
\begin{tabular}{ccccccc}
\hline
\multicolumn{1}{c}{T} &
\multicolumn{1}{c}{$n_B$} &
\multicolumn{1}{c}{$\epsilon_{\bar\nu}^{d}(\delta\mu)$} &
\multicolumn{1}{c}{$\epsilon_{\bar\nu}^{d}(-\delta\mu)$} &
\multicolumn{1}{c}{$\epsilon_{\bar\nu}^{s}(\delta\mu)$} &
\multicolumn{1}{c}{$\epsilon_{\bar\nu}^{s}(-\delta\mu)$} &
\multicolumn{1}{c}{$\delta\mu$}\\
\hline
\multicolumn{1}{c}{($MeV$)} &
\multicolumn{1}{c}{$(fm^{-3})$} &
\multicolumn{1}{c}{$(erg/cm^{3}/s)$} &
\multicolumn{1}{c}{$(erg/cm^{3}/s)$} &
\multicolumn{1}{c}{$(erg/cm^{3}/s)$} &
\multicolumn{1}{c}{$(erg/cm^{3}/s)$} &
\multicolumn{1}{c}{$MeV$} \\
\hline
   &0.8&1.54$\times 10^{+24}$&4.46$\times 10^{+23}$&6.47$\times
10^{+26}$&8.52$\times 10^{+23}$& 1.00 \\
    &1.0&8.31$\times 10^{+23}$&1.78$\times 10^{+23}$&4.20$\times
10^{+26}$&6.51$\times 10^{+23}$&  \\
\cline{2-7}
   &0.8&2.55$\times 10^{+24}$&2.01$\times 10^{+23}$&5.43$\times
10^{+27}$&1.26$\times 10^{+22}$& 2.00 \\
    &1.0&1.51$\times 10^{+24}$&6.51$\times 10^{+22}$&3.00$\times
10^{+27}$&9.87$\times 10^{+21}$&  \\
\cline{2-7}
0.2 &0.8&3.75$\times 10^{+24}$&7.76$\times 10^{+22}$&2.43$\times
10^{+28}$&1.23$\times 10^{+20}$& 3.00 \\
    &1.0&2.53$\times 10^{+24}$&1.82$\times 10^{+22}$&1.07$\times
10^{+28}$&9.75$\times 10^{+19}$&  \\
\cline{2-7}
   &0.8&8.72$\times 10^{+23}$&8.72$\times 10^{+23}$&3.43$\times
10^{+25}$&3.43$\times 10^{+25}$& 0.00 \\
    &1.0&4.10$\times 10^{+23}$&4.10$\times 10^{+23}$&2.49$\times
10^{+25}$&2.49$\times 10^{+25}$&  \\
\cline{2-7}
\hline
   &0.8&3.02$\times 10^{+25}$&1.20$\times 10^{+25}$&6.18$\times
10^{+27}$&2.64$\times 10^{+26}$& 1.00 \\
    &1.0&1.94$\times 10^{+25}$&6.93$\times 10^{+24}$&3.20$\times
10^{+27}$&1.58$\times 10^{+26}$&  \\
\cline{2-7}
   &0.8&4.50$\times 10^{+25}$&7.08$\times 10^{+24}$&2.07$\times
10^{+28}$&4.08$\times 10^{+25}$& 2.00 \\
    &1.0&3.06$\times 10^{+25}$&3.75$\times 10^{+24}$&9.73$\times
10^{+27}$&2.49$\times 10^{+25}$&  \\
\cline{2-7}
0.4 &0.8&6.49$\times 10^{+25}$&3.93$\times 10^{+24}$&5.49$\times
10^{+28}$&5.45$\times 10^{+24}$& 3.00 \\
    &1.0&4.60$\times 10^{+25}$&1.80$\times 10^{+24}$&2.33$\times
10^{+28}$&3.41$\times 10^{+24}$&  \\
\cline{2-7}
   &0.8&1.93$\times 10^{+25}$&1.93$\times 10^{+25}$&1.43$\times
10^{+27}$&1.43$\times 10^{+27}$& 0.00 \\
    &1.0&1.19$\times 10^{+25}$&1.19$\times 10^{+25}$&8.15$\times
10^{+26}$&8.15$\times 10^{+26}$&  \\
\cline{2-7}
\hline
\end{tabular}
\end{center}
\end{table}
\par
For the two flavour quark matter, the anti-neutrino emissivity
increases while that of the neutrino emissivity decreases on increasing
 $\delta\mu$ (same as increasing $-\delta\mu$ in the case of
neutrino). For a constant temperature of $T=0.4~MeV$, the
anti-neutrino emissivity depends on its exponent as
$\varepsilon_{\bar\nu} \propto T^{3.02}$ to $T^{1.47}$  and on that of
the baryon density as $\varepsilon_{\bar\nu} \propto n_{B}^{1.1}$  to
$n_{B}^{1.22}$ for $\delta\mu$ varies from  1. to 2. $MeV$ (0.005 to
0.01$fm^{-1}$) In a similar way, the neutrino emissivity is
proportional to the temperature as well as the baryon density
exponents (for the constant temperature of $T=0.4~MeV$ as in the
earlier case) as $\varepsilon_{\nu} \propto T^{9.72}$ to $T^{16.92}$ and
$n_{B}^{1.03}$ to $n_{B}^{0.84}$ for $\delta \mu$ = -1. to -2.$MeV$
(0.005 to 0.01 $fm^-1$). But for the equilibrium case
($\delta\mu=0$), ${\varepsilon_{\bar\nu}}_{\nu} \propto
T^{5.3 - 5.9}$ and $\propto n_{B}^{1.03}$ as has been observed
earlier \c{154a}.
\par
In addition, for the three flavour quark matter, the behaviour of the
anti-neutrino as well as the neutrino emissivity
is qualitatively same due to the variation in
$\delta\mu$(-$\delta\mu$ in the case of neutrino), as is seen above
in the case of two flavour quark matter. But unlike the two flavour
quark matter, the anti-neutrino and the neutrino
emissivity is inversely proportional to the baryon density.
The variation in the exponents here (for $T=0.4~MeV$ in both cases)
is, $\varepsilon_{\bar\nu} \propto T^{4.55}$ to $T^{4.19}$  and
$n_{B}^{-1.96}$  to
$n_{B}^{-2.16}$ for the $d$ decay, and $\propto T^{2.93}$  to
$T^{1.12}$; and $n_{B}^{-2.95}$  to $n_{B}^{-3.85}$ for the $s$ decay when
one changes $\delta\mu$ from  1. to 2. $MeV$. Similarly,
$\varepsilon_{\nu} \propto T^{5.29}$ to $T^{6.88}$; and $n_{B}^{-2.48}$
to $n_{B}^{-3.51}$  for the $d$ decay, and $\propto T^{7.92}$ to
$T^{15.09}$; and $n_{B}^{-2.31}$  to
$n_{B}^{-2.11}$ for the $s$ decay due to variation in $\delta \mu$= --1.
to --2. $MeV$.
But the results in the equilibrium case ($\delta\mu=0$) which has
already been shown in the literature \c{154a} is
${\varepsilon_{\bar\nu}}_{\nu} \propto
T^{4.86-4.47}$; $\propto n_{B}^{-2.16}$  for the case of $d$ decay
and $\propto T^{5.03-5.38}$, $n_{B}^{-2.53}$
for $s$ decay with $m_{s}=150$ $MeV$ and $\alpha_{c}=0.1$.
\vfill
\par
Therefore, it is obvious from the above numerical results that the
exponent of the temperature decreases but that of the baryon density
increases in the anti-neutrino emissivity case compared to the
equilibrium condition. But for the case of neutrino emissivity, the
exponent of the temperature increases  and the baryon density
exponent decreases. Thus it is clear that the changes on the neutrino
emissivity as well as the anti-neutrino emissivity  is due to the
non-equilibrium condition where $\delta\mu\ne 0$.
The change in the emissivity ($\varepsilon_{\bar\nu} (\varepsilon_{\nu})$)
due to the variation in $\delta\mu(-\delta\mu)$ for $T=0.4~ MeV$ is
shown in Table 8.3 and 8.4 for two flavour and three flavour case, which
has also been extended to $T=0.5~ MeV$ (Table 8.3 and 8.4). Moreover, it
is seen from Table 8.5 and 8.6 that the difference in the chemical
potential affects the density distribution of particles for both two
and three flavour quark matter case. So, when $\delta\mu(-\delta\mu)$
increases from 0.0 to 0.01 $fm^{-1}$, the
Fermi momentum of $u$ quark and $e$ increase (decrease), and that of $d$ quark
decreases (increases) with respect to the chemical equilibrium Fermi
momenta of $u$, $d$ and $e$ for two flavour quark matter. But for the three
flavour quark matter case (for $d$ and $s$ decay), the chemical
equilibrium Fermi momentum of $u$ quark remains the same, and that of $d$
quark and $e$  decrease (increase) and $s$ increases (decreases) as one
changes $\delta\mu(-\delta\mu)$ from 0.0 to 2.0 $MeV$.
Hence, in comparison to the equilibrium case, the neutrino emissivity
decreases monotonically, whereas that of the anti-neutrino
emissivity increases monotonically for both the two and the three
flavour quark matter.

\begin{table}
\cl{\bf TABLE 8.5}
\vspace{0.2in}
\cl{ BARYON NUMBER DENSITY $n_B$, FERMI
MOMENTA OF $u$-QUARK}
\cl{ $p_{F}(u)$, $d$-QUARK $p_{F}(d)$ AND ELECTRON $p_{F}(e)$.}
\vspace {0.2in}
\begin{center}
\begin{tabular}{ccccc}
\hline
\multicolumn{1}{c}{$\delta\mu$} &
\multicolumn{1}{c}{$n_B$} &
\multicolumn{1}{c}{$p_{F}(u)$} &
\multicolumn{1}{c}{$p_{F}(d)$} &
\multicolumn{1}{c}{$p_{F}(e)$} \\
\multicolumn{1}{c}{$fm^{-1}$}&
\multicolumn{1}{c}{($fm^{-3}$)} &
\multicolumn{1}{c}{($MeV$)} &
\multicolumn{1}{c}{($MeV$)} &
\multicolumn{1}{c}{($MeV$)} \\
\hline
   &0.8&393.81&494.40&109.13 \\
0.0&1.0&424.22&532.58&117.56 \\
\hline
   &0.8&393.89&494.35&112.16\\
0.0075&1.0&424.30&532.53&120.59\\
\hline
   &0.8&393.92&494.34&113.17\\
0.01&1.0&424.33&532.51&121.60\\
\hline
   &0.8&393.74&494.45&106.09\\
-.0075&1.0&424.14&532.63&114.52\\
\hline
   &0.8&393.71&494.47&105.08\\
-.01&1.0&424.12&532.64&113.50\\
\hline
\end{tabular}
\end{center}
\end{table}
\begin{table}
\cl{\bf TABLE 8.6}
\vspace{0.2in}
\cl{ BARYON NUMBER DENSITY $n_B$, FERMI MOMENTA OF
$u$-QUARK $p_{F}(u)$, $d$-QUARK}
\cl{ $p_{F}(d)$, $s$-QUARK $p_{F}(s)$ AND ELECTRON $p_{F}(e)$ FOR
$m_s = 150~MeV$ AND  $\alpha_c =0.1$. }
\vspace {0.2in}
\begin{center}
\begin{tabular}{cccccc}
\hline
\multicolumn{1}{c}{$\delta\mu$} &
\multicolumn{1}{c}{$n_B$} &
\multicolumn{1}{c}{$p_{F}(u)$} &
\multicolumn{1}{c}{$p_{F}(d)$} &
\multicolumn{1}{c}{$p_{F}(s)$} &
\multicolumn{1}{c}{$p_{F}(e)$}\\
\multicolumn{1}{c}{($MeV$)} &
\multicolumn{1}{c}{($fm^{-3}$)} &
\multicolumn{1}{c}{($MeV$)} &
\multicolumn{1}{c}{($MeV$)} &
\multicolumn{1}{c}{($MeV$)} &
\multicolumn{1}{c}{($MeV$)}\\
\hline
     &0.8&392.92&395.08&390.73&2.35\\
0.0  &0.1&423.26&424.81&421.69&1.69\\
\hline
     &0.8&392.92&394.62&391.20&1.85\\
1.0  &0.1&423.26&424.35&422.16&1.18\\
\hline
     &0.8&392.92&394.15&391.68&1.34\\
2.0  &0.1&423.26&423.88&422.63&0.68\\
\hline
     &0.8&392.92&395.55&390.25&2.86\\
-1.0  &0.1&423.26&425.28&421.21&2.19\\
\hline
     &0.8&392.92&396.01&389.77&3.36\\
-2.0  &0.1&423.26&425.74&420.74&2.70\\
\hline
\end{tabular}
\end{center}
\end{table}
\vfill
\newpage
\setcounter{equation}{0}
\chapter {Summary and Conclusion}
\hspace{0.3in} In this thesis, we have investigated the
behaviour of matter at very high
density using a relativistic Lagrangian description. The Lagrangian chosen
by us corresponds to the chiral sigma model. This approach is considered
to be a ``good" low energy limit of quantum chromo dynamics. Although
there have been a few
previous calculations along this line, a detailed and consistent field
theoretical approach has been lacking. The calculations presented in this
thesis are aimed at such a detailed study. We have extended these
calculations for the case of finite temperatures ($\leq 15~MeV$). The
results are expected to find application in stellar collapse calculations.
In addition, we have dealt with the following subjects : (1) Phase
transition to quark matter and the possible formation of strangelets
at high densities and (2) astrophysical
applications of our results, to (a) structure and radial oscillation
of nonrotating neutron stars and  (b) the neutrino emissivity
of quark matter with an improved calculation of phase space
integrals involved.
\par
The highlights and main results of this thesis can be summarized
as follows :
\begin{enumerate}

\item  The energy per nucleon of cold nuclear matter ($k_BT=0$), derived by
us using chiral sigma model, is in good agreement with the preliminary
estimates inferred from heavy-ion collision data \c{56} in the
density range between one to four times the nuclear saturation
density ($n_s$).

\item  For a system of high density nuclear matter, based on the
chiral sigma model, we find that a strict first order phase transition to
($u,~d,~s$) quark matter is not favoured. This does not, of course,
preclude a phase transition of second order. However, we have not
investigated the latter problem.

\item The mass formulae for finite lumps of strange quark matter with
$u,~d$ and $s$ quarks and non-strange quark matter ($u$ and $d$) are
derived in a non-relativistic approach, taking into account the finite
size effects such as surface and curvature. We find that there is a
good possibility for the formation of metastable strangelets of large
mass detectable in experiment. This is important since the detection
of strangelets may be the most unambiguous way to confirm the
formation of quark-gluon plasma in heavy ion collision experiment.

\item The maximum mass for stable neutron stars predicted by our
equation of state for ($n,~p,~e$) matter is 2.59 times the solar
mass. The corresponding radius ($R$), crustal length ($\Delta$)
and surface red
shift ratio ($\alpha$) are 14.03 $km$, 1.0 $km$ and 0.674
respectively. The maximum moment of inertia is $4.79 \times
10^{45}~g~cm^{2}$. These suggest that our equation of state for
neutron star matter is comparatively ``stiff". This is reflected in
the value of the maximum mass of neutron stars, which is the largest
for the present model as compared to other available field
theoretical equation of state models. It may be mentioned here that
observational evidence in favour of a stiff equation of state comes
from the identification by Tr\"umper $et~al.$ \c{64} of the 35 day cycle
of the pulsating X-ray source Her X-1 as originating in free
precession of the rotating neutron star (Pines \c{65} for a
discussion). For a 1.4 times the solar mass neutron star
configuration, we get : $R$ = 14.77 $km$, $I$ = 2.15 $\times$
10$^{45}$ $g~cm^2$, $\Delta$=3.0 $km$ and the  red shift ratio (at
the surface) $\alpha$ = 0.85. The corresponding central density is
4.06 $\times$ 10$^{14}$ $g~cm^{-3}$.

\item  The neutrino emissivity from two and three flavour quark matter is
numerically calculated and compared with the result given by
Iwamoto \c{144}. We find that the emissivity is smaller than
Iwamoto's result by about two orders of magnitude when
$p_{f}(u)+p_{f}(e)-p_{f}(d(s))$ is comparable to the
temperature. We attribute this to the
severe restriction imposed by momentum conservation on the phase space
integral. An alternative formula for the neutrino emissivity,
which is valid when the quarks and electrons are degenerate and any
values of $p_{f}(u)+p_{f}(e)-p_{f}(d(s))$ is obtained by us.
\end{enumerate}

\vfill
\end{document}